# Physics and Chemistry on the Surface of Cosmic Dust Grains: A Laboratory View

Alexey Potapov[1] and Martin McCoustra[2]

[1]Laboratory Astrophysics Group of the Max Planck Institute for Astronomy

at the Friedrich Schiller University Jena, Germany

*alexey.potapov@uni-jena.de*

[2]*Institute of Chemical Sciences, Heriot-Watt University, Edinburgh, UK.*

*M.R.S.McCoustra@hw.ac.uk*

Dust grains play a central role in the physics and chemistry of cosmic environments. They influence the optical and thermal properties of the medium due to their interaction with stellar radiation; provide surfaces for the chemical reactions that are responsible for the synthesis of a significant fraction of key astronomical molecules; and they are building blocks of pebbles, comets, asteroids, planetesimals, and planets. In this paper, we review experimental studies of physical and chemical processes, such as adsorption, desorption, diffusion, and reactions forming molecules, on the surface of reliable cosmic dust grain analogues as related to processes in diffuse, translucent, and dense interstellar clouds, protostellar envelopes, planet-forming disks, and planetary atmospheres. The information that such experiments reveal should be flexible enough to be used in many different environments. In addition, we provide a forward look discussing new ideas, experimental approaches, and research directions.

**Review Contents**

*"We come from stardust and to stardust we shall return"* (Joan Oró).

## 1. A Little Bit of History

By the late 1960s, it had become clear that chemical processes occurring on the surfaces of dust grains were likely responsible for the formation of hydrogen ($H_2$) and other small hydrides in interstellar environments [1]. However, it was not until the 1990s that the surface science community began to interact closely with the astronomy and astrophysics communities in seeking to understand these processes in detail. The early discussion between these communities was focussed on the key question, from the surface science standpoint, of the nature of the surfaces that should be investigated. The observational data pointed to what was likely to be the complex nature of grain materials that might be found in space. That complexity is, in part, revealed through images of interplanetary dust particles (IDPs) such as in **Figure 1** [2].

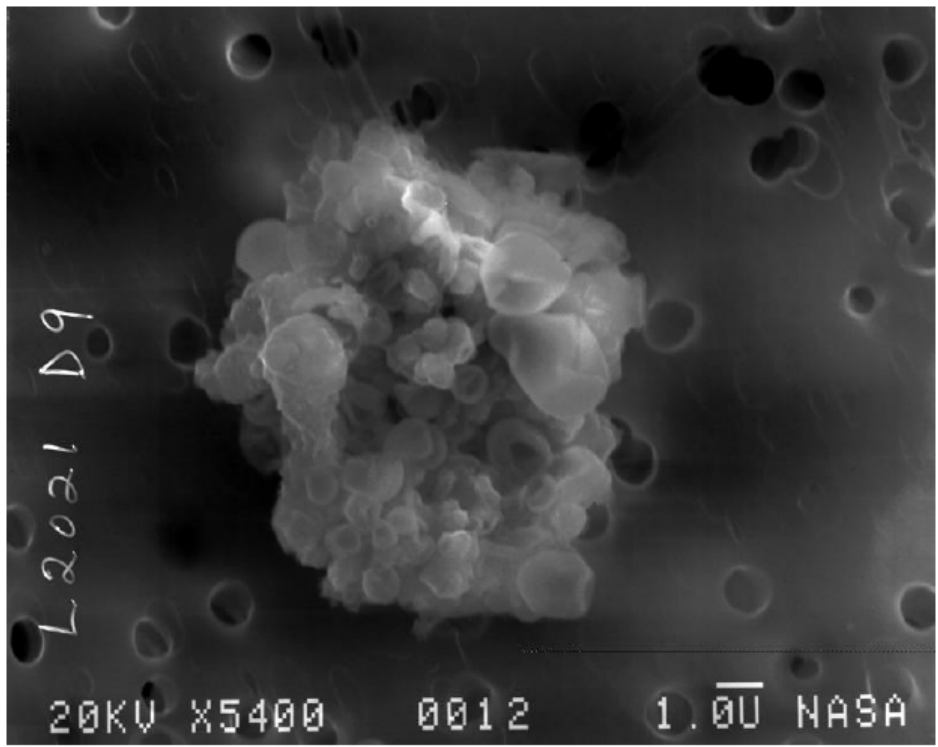

**Figure 1.** An electron microscope image of the interplanetary dust particle (IDP) L2021D9 showing its agglomerate nature. Reproduced from NASA catalog [2].

However, from a surface science standpoint such compositionally variable and fractal materials were not considered as suitable simple, consistent and reproducible models on which to build an understanding of the surface science of grains. Hence, by the mid-1990s, the surface science community had lighted upon two substrate options; graphite (as highly oriented pyrolytic graphite, HOPG) and silica / silicate minerals as either bulk crystalline solids or amorphous nanoparticle thin films on metal substrates. The application of surface science to developing our understanding of the interstellar gas-grain interaction grew from there.

## 2. Formation and Role of Dust in Cosmic Environments

Matter in the Universe evolves through the so-called *cosmic life cycle of matter*. The word "interstellar" is attached to this matter as it describes its affiliation to the interstellar medium (ISM) – the space between the stars with surrounding gravitationally attached material. The cosmic life cycle can be thought to begin with the diffuse interstellar medium, where number densities, $n$, are typically 50 cm$^{-3}$ and temperatures are around 80 K, which is dominated by neutral hydrogen atoms. The diffuse ISM evolves into translucent clouds and then into cold, dense interstellar molecular clouds (which may or may not contain prestellar cores) with a typical $n$ of $10^4$ cm$^{-3}$ and temperatures as low as 10 K, where molecular hydrogen and many other molecular species are found. Dense clouds collapse into denser cores forming central stars surrounded by protostellar envelopes that evolve into protoplanetary (or planet-forming) disks having large number density gradients ranging from $10^4$ to $10^{12}$ cm$^{-3}$ and temperature gradients from 10 to several 1000 K. Planet-forming disks evolve into planetary systems containing planets and debris disks (remnants of protoplanetary disks after planet formation). Central stars at the end of their life (so-called evolved stars) expand absorbing the circumstellar material and, finally, release their shells into the ISM feeding the diffuse interstellar medium. For detailed descriptions of the stages of the cosmic life cycle of matter, we refer the reader to a number of review papers [3-6] and books [7,8].

Dust is one of the fundamental players in the physical and chemical processes taking place in the cosmic environments mentioned above. Beginning its lifecycle in the envelopes of evolved stars and, in their later evolutionary stages, planetary nebulae and supernovae; dust enters the ISM, where it can be destroyed and re-condensed; and travels through diffuse and dense clouds, protostellar envelopes, and planet-forming disks. Dust influences the thermodynamic properties of the medium by absorption and emission of stellar radiation; serves as a third body and a catalyst for key chemical reactions; participates in the formation of planets and their atmospheres and, possibly, in the delivery of the chemical precursors to life to those planets; and ends its lifecycle in the circumstellar media of dying stars.

Carbonaceous and siliceous materials are the major components of the interstellar dust. Other types of solids, such as metallic oxides, carbides, sulfides and even metals themselves, are minor components [9-13]. The formation pathways of interstellar dust remain a matter of significant scientific discussion. Primary dust formation occurs *via* gas-phase reactions in high-temperature circumstellar envelopes of evolved stars, mainly of asymptotic giant branch (AGB) stars. AGB stars have extended atmospheres rich in oxygen, silicon, and carbon. These elements are manufactured in the stellar core and are dredged up to the surface by convection currents in the hot stellar plasma. In oxygen-rich AGB stars, where $n(O)/n(C) > 1$ and carbon is predominantly trapped in carbon monoxide (CO), oxygen atoms react in the stellar atmosphere with silicon and any other metal atoms to form amorphous and crystalline oxide and silicate grains. Carbon-rich stars, where $n(O)/n(C) < 1$, give rise to carbon particles in the form of graphite or amorphous carbon grains. These primary dust particles or grains have sizes from a few to tens of nanometres [14,15].

After formation in circumstellar shells of evolved stars, dust grains are blown away from their parent stars by radiation pressure into the ISM, where they can be completely destroyed by supernova shocks. Estimates show that only a part of the dust produced by stars survives into the ISM [16-18]. However, observations demonstrate the presence of dust grains in high concentrations in the ISM. Thus, an efficient formation mechanism is required to explain the difference between stellar formation and interstellar destruction rates. This alternative mechanism for dust grain formation and growth may be cold condensation under interstellar conditions. Laboratory experiments have demonstrated efficient formation of nanometre-sized silicate and carbon grains at low temperatures [19-21]. In this process, hydrogen molecules may play a moderating role [22].

In the ISM, dust grain temperatures drop to as low as 10 - 20 K as the density increases; and they can acquire additional hydrogen, oxygen, carbon, nitrogen, and sulfur atoms, which have also escaped from stars. Here, intense dust growth takes place, mainly defined by two mechanisms; the addition of gas-phase atoms and molecules, and coagulation *via* grain-grain collisions. Consequently, dust grains in the ISM reach micrometre-sizes [10]. In addition, molecule formation on dust grain surfaces takes place, with mobile atoms travelling across the surfaces to meet each other and react giving birth to basic astrophysical molecules such as $H_2$ and $H_2O$.

In denser regions of the ISM, *molecular clouds*, small molecules, such as CO and $N_2$, adsorb on grain surfaces and, together with surface synthesised molecules, build up icy mantles. These mantles include $H_2O$ (the main constituent accounting for more than 60% of the ice in most lines-of-sight [23]), CO, $CO_2$, $NH_3$, $CH_4$, $CH_3OH$ and other minor components [3,24]. Rich chemistry takes place in and on these icy mantles in dense, cold environments (molecular clouds, protostellar envelopes and planet-forming disks beyond the snowline) which ultimately results in the formation of complex organic molecules (COMs), *e.g.* alcohols, amines, aldehydes and ketones, which are considered as precursors to prebiotic species. Astrophysically, COMs are C-bearing molecules consisting of 6 or more atoms. So, *e.g.*, methanol ($CH_3OH$) is a COM from the astrophysical/astrochemical point of view. This is quite different from what is understood by "complex molecules" in chemistry and biology. Prebiotic molecules have not yet been unambiguously detected in the ISM, protostellar envelopes or planet-forming disks; but many precursors have been identified. However, amino acids and other key biological molecule precursors such as simple sugars, purines and pyrimidines have been detected in comets and meteorites [25-28]. Amino acids are the building blocks of proteins and, thus, are critical to life on Earth. Results of laboratory experiments mimicking the physico-chemical processes in the ISM show that the generation of amino acids in the ISM is possible [29-33]; supporting the hypothesis of their formation in the ISM and subsequent exogenous delivery to Earth *via* comets, asteroids and their meteoritic remains [27,34-36].

In protostellar envelopes and planet-forming disks (circumstellar media of young stars), grains continue to grow reaching millimetre sizes [37]. In disks, grains are believed to stick together, held so by molecular forces and form kilometre-sized planetesimals that, in turn, accrete and form planets due to their gravity. Another mechanism driving planetary growth linked to the observational detection of large amounts of pebbles in protoplanetary disks and supported by theoretical calculations is considered to be pebble accretion revolving around millimetre-to-centimetre-sized pebbles [38]. Chemistry in envelopes and disks continues to be very complex [4]. The high densities and temperatures; and intense UV irradiation from a central star make them, possibly, more appropriate environments for the formation of prebiotic molecules than molecular clouds. In reference to the formation of planets, in particular extrasolar planets, dust grains play an important role in the formation of not only planetary bodies but also planetary atmospheres defining their dynamics and chemistry [39]. The formation of planets is also often accompanied by the formation of debris disks around a central star [40]. In debris disks, small dust grains can be formed top-down through collision cascades started by collisions of planetesimals.

Adsorption and desorption are the two processes responsible for linking the gas-phase to the solid state and *vice versa*. The dust surface influences (or in some cases defines) the efficiencies of these processes. Adsorbed and desorbed molecules correspondingly contribute to the physics and chemistry in the solid state and in the gas phase; and their presence defines the physical (pressure, temperature, opacity) and chemical (composition, concentration, reaction networks, reaction rates) properties of the environment.

In all environments discussed above, dust grains absorb and scatter stellar radiation and re-emit the absorbed energy at longer wavelengths. About 30 - 50% of the starlight emitted by our Galaxy is absorbed by the dust. This energy is re-radiated as the far-infrared continuum emission. Optical properties (absorption, emission and scattering) of grains in different spectral regions are important for modelling and understanding the physics of these environments. The opacity of grains underpins estimates of important astrophysical parameters, such as dust temperatures and compositions, mass loss rates of evolved stars, and the total dust mass in interstellar and circumstellar media. Due to the absorption of light, dust grains contribute to the creation of the visible Galaxy. **Figure 2** shows the whole sky as visible to the human eye. In Jules Verne's book "The Children of Captain Grant", one of the heroes looks at the sky and sees a lot of stars and "black holes". Now we know that these "black holes" are molecular clouds that obscure visible light from background stars due to the large concentration of gas and dust.

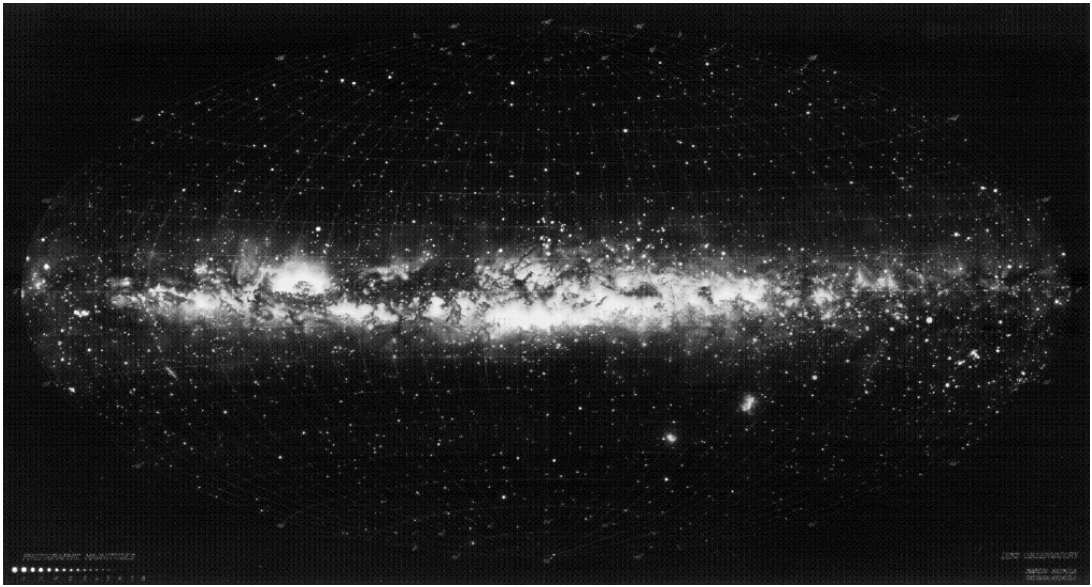

**Figure 2.** The whole sky as visible to the human eye. Image credit: Lund Observatory, Sweden.

## 3. Laboratory Astrophysics

Laboratory astrophysics bridges studies of atomic and molecular species in interstellar and circumstellar media and planetary atmospheres conducted through astronomical observations, and studies of these species or their analogues that we can create in the laboratory and probe *in* *situ*. Laboratory astrophysics provides spectroscopic fingerprints of species that are necessary for decoding astronomical spectra and allows us to study physico-chemical processes in the conditions of cosmic environments. It helps to more completely understand these processes and to build reliable astrochemical models that, being supported by astronomical observations, can reveal our astrochemical past and future.

There are basically three reasons for acquiring physicochemical laboratory data. The first reason is the most obvious and that is getting spectral signatures that can be used to remotely identify and quantify species. In some cases, *e.g.*, CO, $NH_3$, and their isotopologues, these species can be used as local environmental probes (temperature and density). But for the most part, laboratory spectral signatures are necessary to identify the rich distribution of species found in a particular object. This holds true for the solid state too in terms of identifying solid-phase materials, such as dust grains and ices. The second reason is subtler and relates to why kinetic simulations are done at all. These simulations are designed mainly to inform about the chemical evolution of environments and therefore need chemical reaction rate coefficients and their temperature dependences; photolysis and radiolysis rates and their energy dependences; adsorption and desorption rates and their temperature dependences; and reaction branching ratios. Such simulations are using the chemistry as a clock enabling estimates of object age to be made from comparison of models with observation. The final reason for laboratory data is to try to understand the complexity of chemistry that can be generated abiotically and so addressing the question as to what is the nature of the base organic input into a proto-evolutionary system on a nascent planet.

Dust grains in cold dense astrophysical environments (on which we mainly focus in this review), such as molecular clouds, protostellar envelopes and planet-forming disks beyond the snowline, are typically considered to consist of dust particles and molecular ices. Many of the laboratory experiments modelling physical and chemical processes on the surface of cosmic dust grains have been performed on ices covering standard laboratory substrates, such as gold, copper or potassium bromide (KBr), which are not characteristic of cosmic dust grains. This is due to the typical view of dust as comprising a compact core surrounded by a thick ice mantle. Physical and chemical processes occurring on and in such ices are considered independent of the properties of the dust surface. Adsorption, desorption, and reactivity of different molecules and radicals on and in such ices have been studied extensively since the late 1990s. We refer the reader to a number of review papers on these topics [41-46].

However, recent laboratory studies have shown that the bare dust material would be available for the surface processes in cold astrophysical environments. The experiments on the agglomeration of $H_2O$ molecules on the dust surface by Rosu-Finsen *et al.* and Marchione *et al.* [47,48] and on the thermal desorption of $H_2O$ and CO ices mixed with dust grains by Potapov *et al.* [49,50] clearly demonstrate this finding. Cosmic dust grains may be covered by a sub-monolayer or few layer quantities of ice due to their fractal nature, high porosity and corresponding large surface area (see ref. [50] for a discussion). In addition, agglomeration of $H_2O$ molecules may result in grain surfaces presenting both wet and dry areas even for a "thick" ice mantle. In **Figure 3**, we present a sketch showing porous cosmic dust grains covered by ice molecules.

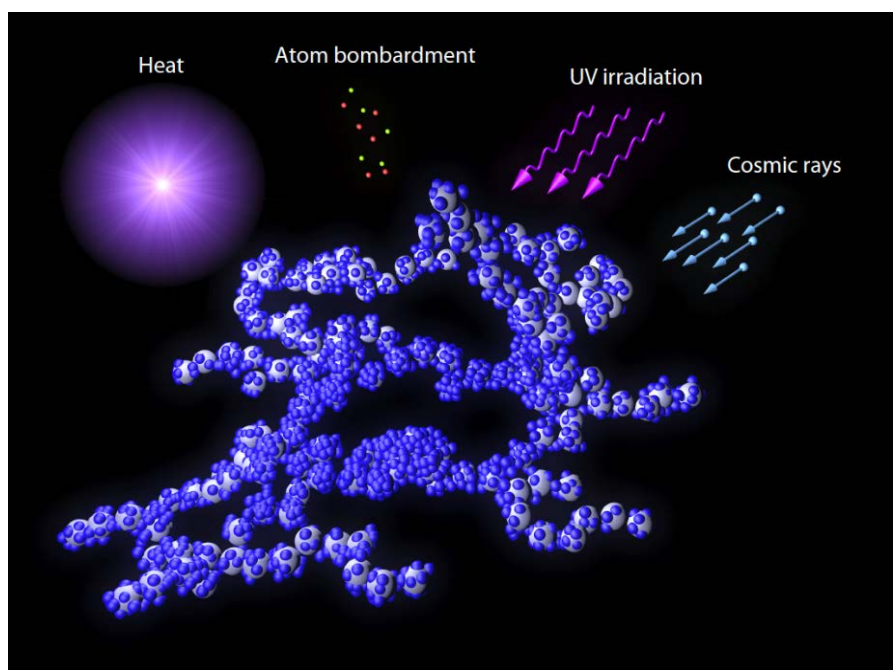

**Figure 3.** Schematic figure showing dust grains (in grey) mixed with ice (in blue) and the main sources of their processing in astrophysical environments. Reproduced from [50].

Thus, the role of the dust surfaces in physical and chemical processes has been underestimated and the studies discussed in this review are of great relevance to understanding the real physico-chemical picture of cold cosmic environments. Summarising the findings of the laboratory studies of the physics and chemistry on the surface of reliable dust grain analogues (amorphous carbon grains, atomic carbon foils, graphite, amorphous silica, and amorphous and crystalline silicate grains), it has been shown that: (i) the efficiency of molecule formation depends on the morphology of the grain surface; (ii) the binding energies of species can be quite different for different grain surfaces and on grain surfaces compared to ice surfaces; (iii) functional groups and atoms in the grain surface can participate directly in surface reactions; (iv) the grain surface has a catalytic effect; (v) desorption kinetics and yields of volatile molecules are different for different grain surfaces, ices and mixtures thereof; (vi) water molecules can be trapped on the grain surface at temperatures beyond the desorption temperature of water ice. Dust grains themselves therefore play an important role in the processes occurring on their surfaces. Thus, our main purpose for this review is to direct the attention of the astrophysical, astrochemical, and astronomical communities to the importance of laboratory experiments and models dealing with surface processes on reliable cosmic dust grain analogues.

## 4. Origins, Composition and Structure of Cosmic Dust
### 4.1. Origins of Dust

Interstellar dust grains comprise only around 1% of the mass of the ISM, or about 0.1% of the Galactic mass, but play a fundamental role in its evolution. Most of the dust found in space comes from AGB stars which have moved off the main sequence and entered the red giant phase of their evolution. As outlined in Section 2, these stars have extended atmospheres with variable composition but are basically either O-rich, producing duty silicate and oxide materials, or C-rich, from which sooty carbon-based materials are derived. More massive silicon-burning stars probably end their lives as supernovae ejecting precursor molecules (*e.g.* SiO) and silicate dust into the ISM. Radiation pressure from the star or supernova pushes the stellar dust out into the depths of interstellar space where it evolves (and, probably, re-forms) in response to the interstellar radiation and particle fields, supernovae shocks and gases to which it is exposed in its journey through space. Such space weathering of dust alters its bulk and surface structural and chemical properties and is a significant independent topic, which we feel is outside the remit of this review. Some examples can be found in recent papers and references therein [51,52]. However, there is definitely room for a review focussing on this topic in detail and describing grain destruction in shocks; low-temperature condensation of dust; grain modification by cosmic rays, UV radiation, heat, and atoms; and related phenomena in terms of observational and laboratory experiments.

### 4.2. Composition of Dust

Stellar spectroscopy across the electromagnetic spectrum has allowed us to establish an elemental composition for the local Galaxy. In the ISM, the observed abundances of heavy elements are found to be depleted, *i.e.* reduced, from the stellar composition [53]. This is consistent with these materials being locked up in the solid phase including dust grains and clearly connects dust to its stellar origins. The depletion is found to scale with the density of the local environment; cold, dense environments show greater depletion than cold, diffuse environments which, in turn, show more depletion than warm diffuse environments. This reflects grain evolution and ice mantle formation on colder grains.

Silicate materials form a considerable fraction of the total mass of interstellar dust. Interstellar silicate grains are likely to contain Fe and Mg, as both of these elements are astrophysically abundant and strongly depleted. At least ~95% of the silicate material is amorphous [54]. The principal forms expected are pyroxenes, $Mg_xFe_{1-x}SiO_3$ ($0 < x < 1$), including enstatite ($MgSiO_3$) and ferrosilite ($FeSiO_3$); and olivines, $Mg_{2x}Fe_{2-2x}SiO_4$, including fayalite ($Fe_2SiO_4$) and forsterite ($Mg_2SiO_4$). Additionally, the pyroxenes exhibit polymorphism as orthopyroxene and clinopyroxene. All of these materials are common in meteorites, and spectral signatures of enstatite and forsterite have been seen in the dusty shells around AGB stars. Olivines have been seen in dust grains captured by the Stardust mission [55]. Silicate grains typically dominate dust emission in many astrophysical environments, and are observed in the cold neutral medium, comets, protoplanetary disks, and even in the far-distant Universe [56]. The 10 μm Si-O stretching feature and the 18 μm O-Si-O bending feature are universally seen in absorption. Though grain formation in stellar outflows likely occurs under high temperature and pressure, near-equilibrium conditions and yields crystalline materials, these spectral features are broad and relatively featureless reflecting extensive space weathering producing amorphous silicate materials.

Crystalline silicates are observed towards young stars which suggest that *in situ* formation by thermal annealing or shocks may have occurred. Some dust in circumstellar shells of evolved stars displays emission features characteristic of crystalline silicates, and from the locations and strengths of these features, it is possible to infer the Mg:Fe ratio. In all examples known so far, the crystalline silicate material appears to be very Mg-rich and Fe-poor, which is consistent with pure forsterite and enstatite. However, Molster *et al*. [57-59] reported that enstatite ($MgSiO_3$) is more abundant than forsterite ($Mg_2SiO_4$) around most evolved stars.

Observed gas-phase abundances of magnesium, iron and silicon have been interpreted to possibly indicate that silicate grains may have a magnesium-rich mantle and iron-rich core [60]. The total ratio of Fe+Mg:Si in grains is estimated to be about 3:1; in excess of the 2:1 ratio expected for olivines, suggesting that metal oxides or metallic Fe may be present as well as silicates [61,62].

Carbonaceous dust grains include pure carbon in both crystalline forms (*i.e.* diamond and graphite) and amorphous or glassy forms (*i.e.* composed of a mixture of graphite and diamonds), alongside hydrocarbons in the form of hydrogenated amorphous carbons, polycyclic aromatic hydrocarbons (PAHs), aliphatic hydrocarbons and fullerenes [12,63]. A nice example of the evolution of carbonaceous materials, PAHs in particular, is reported by Song *et al.* in observations of the Red Rectangle [64]. Also seen, but for probably less than 5% of all forms of carbonaceous grains, are other carbonaceous compounds including silicon carbide (SiC) and carbonates such as calcite ($CaCO_3$) and dolomite ($CaMg(CO_3)_2$) [65].

Micrometre-sized pre-solar grains such as graphite, silicon carbide (SiC), corundum ($Al_2O_3$) and silicon nitride ($Si_3N_4$), and nanometre-sized pre-solar grains (*e.g.*, nano-diamonds and titanium carbide nano-crystals) of interstellar origin as indicated by their anomalous isotropic composition have been identified in primitive meteorites [66]. Pre-solar silica grains have been identified in IDPs [67]. Sub-micrometre-sized GEMS (glass with embedded metals and sulfides) of pre-solar origin have also been identified in IDPs, and their 8-13 μm absorption spectrum is similar to that observed in dense molecular clouds and young stellar objects [68].

**4.3. Structure of Dust**

Models of observed interstellar extinction, between 100 nm and 1 μm, feature generalized particle size distributions of graphite, enstatite, olivine, silicon carbide, iron and magnetite or combinations of these materials. The extinction curve rises from the near-IR to the near-UV with a broad absorption feature at around 217.5 nm consistent with graphitic material, followed by a steep rise into the far-UV at 100 nm. The size distributions are power laws, which monotonically decrease from 0.005 μm towards larger sizes beyond 2.5 μm. The presence of nanometre-sized or smaller particles in the interstellar medium is indicated directly by the interstellar far-UV extinction; the ubiquitous 3.3, 6.2, 7.7, 8.6, and 11.3 μm PAH emission features; the near- and mid-IR broadband emission seen in the IRAS 12 and 25 μm bands and the COBE-DIRBE 3.5, 4.9, 12 and 25 μm bands; the 10-100 GHz Galactic foreground microwave emission; and indirectly by interstellar gas heating [69]. Very large grains, with radii of 1 μm, which enter the solar system from the ISM, have been detected by the Ulysses dust detector [70]. Very large grains with radii of around 10 μm, whose interstellar origin has been indicated by their hyperbolic velocities, have been detected by radar methods [71]. However, Frisch *et al.* [72] and Weingarther and Draine [73] have argued that the volumes of very large grains which have been inferred from the detections were difficult to reconcile with the interstellar extinction and interstellar elemental abundances.

The shape of dust grains is generally approximated to be spherical or cylindrical in order to simplify electromagnetic field calculations. However, dust grains are unlikely to be spherical.

It is more likely that grains stick together, collide and shatter, leading to a rather random distribution of shapes. The shape of the dust grains has a large influence on the shape of the absorption and extinction spectrum of the grains [74]. Grains come in a range of shapes as polarisation in the interstellar radiation field indicates that some fraction of the grains must be non-spherical and aligned. Since the wavelength dependence of the interstellar polarisation exhibits a steep decrease towards the UV, this suggests that the ultra-small grain components, which are responsible for the far-UV extinction curve, are either spherical or unaligned [75]. An understanding of grain shapes is important for working out grain dynamics and for the physics and chemistry of molecule formation on grain surfaces.

Analysis of cometary dust particles, dust evolution models, and laboratory experiments have indicated that grains in interstellar clouds, protostellar envelopes, and protoplanetary disks may be very porous. Dust evolution models showed that grains in the ISM grow to fractal aggregates of sub-micrometre sizes having a porosity of up to 80% [76]. This result was supported by the laboratory experiments on gas-phase condensation of nanometre-sized amorphous carbon or silicate grains and their subsequent deposition onto a substrate, where they aggregate. The structure of grains was described as porous layers of fractal agglomerates having the porosity of up to 90% [77,78]. Fluffy, highly porous aggregates were also produced in the experiments on cold condensation at interstellar conditions [19,20]. Laboratory collision studies [79-81] and models [82-85] relevant to the evolution of grains in planet-forming disks have shown that aggregation of μm-sized grain monomers leads to the formation of fluffy fractal particles with the porosity of more than 90%. Analysis of radar data [86], *in situ* measurements of cometary comae [87-90], modelling of the light-scattering properties of the cometary dust [91], and analysis of the samples collected and returned by the Stardust mission [92] showed that cometary dust particles present a mixture of dense and fluffy, highly porous aggregates. The fluffy particles are typically considered as fractal aggregates of nanometre-sized grains that may be linked to interstellar dust.

## 5. Production and *in situ* Characterisation of Dust Grain Analogues
### 5.1. Production of Dust Grains

Much of the initial application of surface science techniques to the study of the gas-grain interaction focussed, as indicated in **Section 3**, on simple flat surfaces such as noble metals, HOPG and polished mineral disks. Recent years have seen a move to more realistic mimics utilising nanoparticle films as in the laboratories of the authors of this review. Laboratory gas-phase condensation techniques are typically used to mimic astrophysical condensation processes of cosmic dust grains and to produce cosmic dust analogues. Such dust grains can be produced in two steps. The first step is the formation of nanometre-sized grains in the gas phase by laser ablation or pyrolysis techniques. In the second step, the generated grains are deposited as a film on a substrate, where they aggregate forming agglomerates in the size range of up to several tens of nm. For details, we refer the reader to a number of research papers [77,78,93,94] and a review [95]. In this light, it is worth also mentioning the StarDust machine recently developed for the production (*via* atomic gas aggregation), manipulation and *in situ* analysis of nanoparticles [96,97].

The apparatus available to the Laboratory Astrophysics Group of the Max Planck Institute for Astronomy is described in detail elsewhere [77]. In the experiments simulating the formation of cosmic dust, the deposition of nanometre-sized amorphous carbon and silicate particles is performed by pulsed laser ablation of graphite, MgSi, FeSi or MgFeSi targets and subsequent condensation of the evaporated species in a quenching atmosphere of a few mbar $He/H_2$ for carbon grains (addition of $H_2$ allows the production of hydrogenated grains) or $He/O_2$ for silicate grains. The low pressure regime applied is comparable to the pressure conditions for dust condensation in AGB stars [98]. The condensed dust grains are extracted adiabatically from the ablation chamber through a nozzle and then a skimmer decoupling the particles from the carrier gas and generating a particle beam, which is directed into a separate chamber, where the dust grains are deposited onto an IR transparent substrate. The substrate can be cooled down to 10 K allowing simultaneous deposition of dust with ices through an additional gas dosing line. The dust layer on the substrate consists of individual particles in the few nanometre size range and large highly-porous fractal particle aggregates in the size range of up to several tens of nanometres. **Figure 4** shows electron microscopy images of amorphous carbon grains, where high porosity and large surface area of grains are clearly observable.

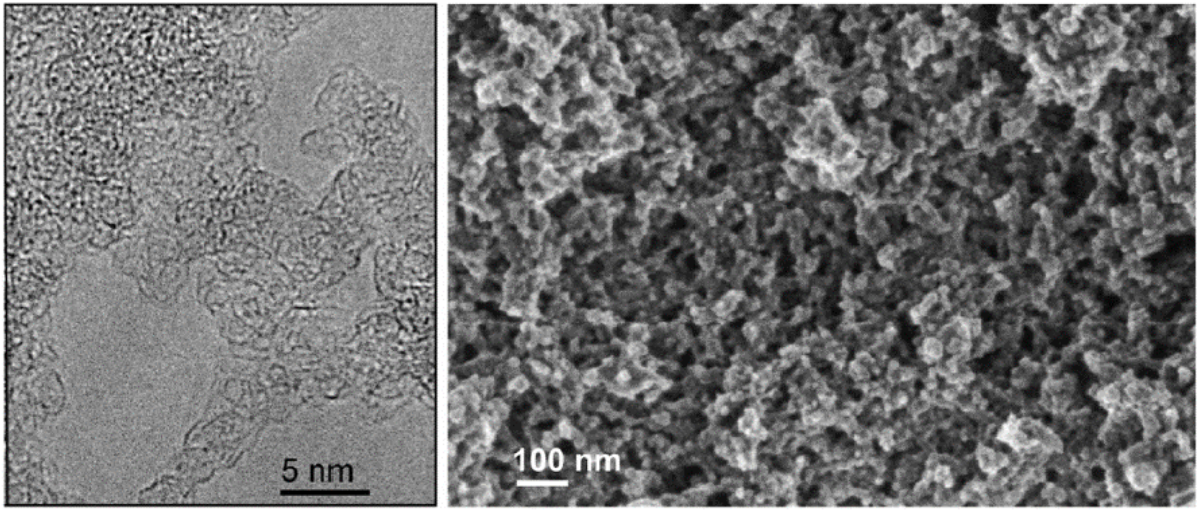

**Figure 4.** Electron microscopy images of amorphous carbon grains produced by gas-phase condensation and subsequent deposition onto a substrate. The left-hand, high-resolution transmission electron microscopy image shows the porous structure of the condensed carbonaceous grains. The agglomerate of nanometre-sized primary grains, which are attached to the lacey carbon support film, is visible in the left upper corner. The right-hand image is a field-emission scanning electron microscopy image of the porous grain layer. Visible individual grains are still agglomerates of smaller grains. Reproduced from [50].

A much simpler approach is taken at Heriot-Watt University. Thin film deposition technologies using electron beam and ion beam evaporation [99,100] are well-established and technologically important in the production of thin film devices and coatings. The former has been adapted to provide for the growth of silica on a polished copper substrate at room temperature [101]. The copper substrate is allowed to oxidise slightly before use to ensure sufficient adhesion of the growing film and the low growth temperature limits mobility during growth with the result that a film comprising silica ($SiO_2$) nanoparticles of a few tens of nanometre in scale are formed. IR spectroscopy reveals the film to be amorphous in nature.

**5.2. Characterisation of Dust Grains**

In studying cosmic dust grain analogues, molecular ices, and dust/ice mixtures, two main *in situ* techniques have come to the fore, infrared (IR) spectroscopy for studying the solid state and mass spectrometry often combined with temperature-programmed desorption (TPD) experiments (uniform sample heating at a constant rate) to probe the gas phase.

Modern IR spectroscopy fully embraces the advantages of interferometry [102] and most, if not all, modern experiments utilise Fourier Transform Infrared (FTIR) instruments to provide identification of species on surfaces and in the solid state through characteristic group vibrational frequencies. Two configurations are in common use. Transmission IR spectroscopy is perhaps the most common and reflects the normal laboratory application. However, it is insufficiently sensitive to be able to identify monolayer and sub-monolayer quantities of materials and so Reflection-absorption Infrared Spectroscopy (RAIRS) has been widely adopted as a result of the 40 to 60-fold enhancement in sensitivity that can be obtained in a near-grazing angle reflection from a reflective metal substrate.

These techniques are complemented by other vibrational spectroscopic techniques, *e.g.*, Raman spectroscopy, and by techniques that access other spectral windows in the ultraviolet-visible (UV-VIS) and terahertz (THz). The reader can find technical details and examples in other review papers (*e.g.*, [41,44,46]). Additionally, surface science has a panoply of methods with which to study surfaces. However, many of these are special and lack generable applicability. X-ray photoelectron spectroscopy (XPS) [103], secondary ion mass spectrometry (SIMS) [104] and low-energy ion scattering spectroscopy (LEIS) [105] have much potential for astrochemistry if properly applied. We refer the reader interested in a wider appreciation of the potential of surface science tools to surface science textbooks (*e.g.*, [106,107]).

There are also many *ex situ* techniques used for characterisation of solid materials. These are outside the scope of this review; though some are discussed in the textbooks mentioned above.

## 6. Chemical Physics on Dust Surfaces

The interaction of atoms and molecules with surfaces is moderated through a potential energy surface (PES). As long ago as the 1930s, Lennard-Jones described, in a one-dimensional manner (**Figure 5**), the nature of the PES for gas-surface interactions in terms that allow us to describe extremes of behaviour [108]: *chemical adsorption* (*chemisorption*) involving the transfer of electrons and the formation of a covalent bond between the surface and adsorbate; and *physical adsorption* (*physisorption*) where the interactions are determined by simple electrostatics and dispersion, *i.e.* van der Waals interactions and hydrogen bonding.

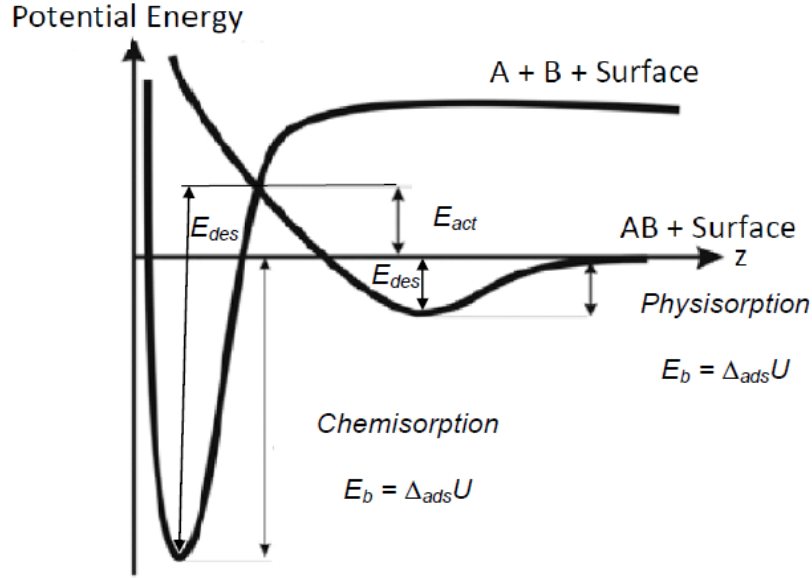

**Figure 5:** Lennard-Jones potential energy surfaces for chemisorption and physisorption adapted by the authors from [109] (Copyright Elsevier). The binding energy $E_b$ is equivalent to the change in internal potential energy between the surface bound and free states, $\Delta_{ads}U$.

Lennard-Jones also identified *activated* and *non-activated* adsorption [108]. The former is typically associated with chemisorption where dissociation of a bond in the adsorbate (*e.g.* dissociation of molecular nitrogen on transition metal surfaces in heterogeneous ammonia synthesis [110]) or restructuring of the surface (*e.g.* in the adsorption of hydrogen atoms to sp$^2$ hybridised carbon materials like PAHs, graphene and graphite [111-113]) is a necessary prerequisite which demands an energetic penalty in the form of an activation energy, $E_{act}$. In such situations,

$$E_{act} = E_{des} - E_b \qquad (1)$$

where $E_{des}$ is the activation energy for desorption and $E_b$ is the binding energy of the adsorbate on the surface. As most systems of astrophysical interest express physisorption behaviour and physisorption is barrier-less, $E_b$ and $E_{des}$ are equal in magnitude but opposite in sign:

$$E_b = -E_{des} \qquad (2)$$

That is not to say that we should ignore chemisorption in astrophysical circumstances. The conditions in astrophysical environments vary significantly. Physisorption dominates in cold, dense environments where the lack of an activation barrier to adsorption ensures that physisorption is efficient even at the lowest of temperatures that we might encounter. In contrast, the activation necessary to produce chemisorption is likely only to be possible in much higher temperature, low density environments where neutral hydrogen atoms dominate.

Though Lennard-Jones presents a simple and appealing model of gas-solid interactions, and is especially attractive when considering physisorption interactions [114], there are clearly issues in this simplicity. In the first instance, if we consider the simplest system of an atom and single crystal surface, then we would require three dimensions to represent the interaction potential; *x* and *y* in the surface plane representing the position within a rigid surface unit cell on a rigid solid and *z*, the perpendicular distance from the surface unit cell. Replacing the atom with a diatomic molecule adds further dimensions to accommodate the rotation and vibration within the molecule [115]. Changing the surface plane will change the potential though state-of-the-art computational surface science using machine learning methods can address this issue and the relationship of single crystal planes to nano- and micro-crystalline catalysts [116]. Allowing for non-rigidity in the surface unit cell adds three dimensions per atom. If the atoms in the underlying solid are considered, the number of dimensions grows significantly. It will grow further if polyatomic molecules are brought into the picture. Amorphous systems add further complexity.

There are numerous approaches to the computation of these potentials for astrochemical purposes. But this is not the place to address this in any detail. Rather it might be hoped that the future will see a suitable review from experts in the field as one currently does not exist.

### 6.1. Adsorption at Surfaces

If we ignore macroscopic mass transport to and from grain surfaces (*e.g.*, mass transport through the pores of grains), the interaction of the gas phase can be described in terms of a balance of fluxes to and from the grain surface as in **Figure 6**.

The input flux, $J_{\text{in}}$, can be equated with the classical wall collision rate, $Z_w$:

$$J_{\text{in}} = Z_w = \frac{P}{\sqrt{2\pi m k_B T}} = \frac{nRT}{\sqrt{2\pi MRT}} \quad (3)$$

where *P* is the partial pressure of a gas of molecular mass, *m*, at an absolute temperature, T; *n* is the equivalent number density and *M* is the molar mass. The efficacy of adsorption can be described by the *sticking coefficient*, *S*, given by

$$S = \frac{J_{\text{ads}}}{J_{\text{in}}} \quad (4)$$

where $J_{\text{ads}}$ is the adsorbed flux. The *accommodation coefficient*, α, defined as

$$\alpha = \frac{J_{\text{ads}} + J_{\text{des}}}{J_{\text{in}}} \qquad (5)$$

is also commonly used to describe uptake at surfaces. The relationship between $S$ and $\alpha$ is clear; as is that $S$ can only be measured in the absence of desorption.

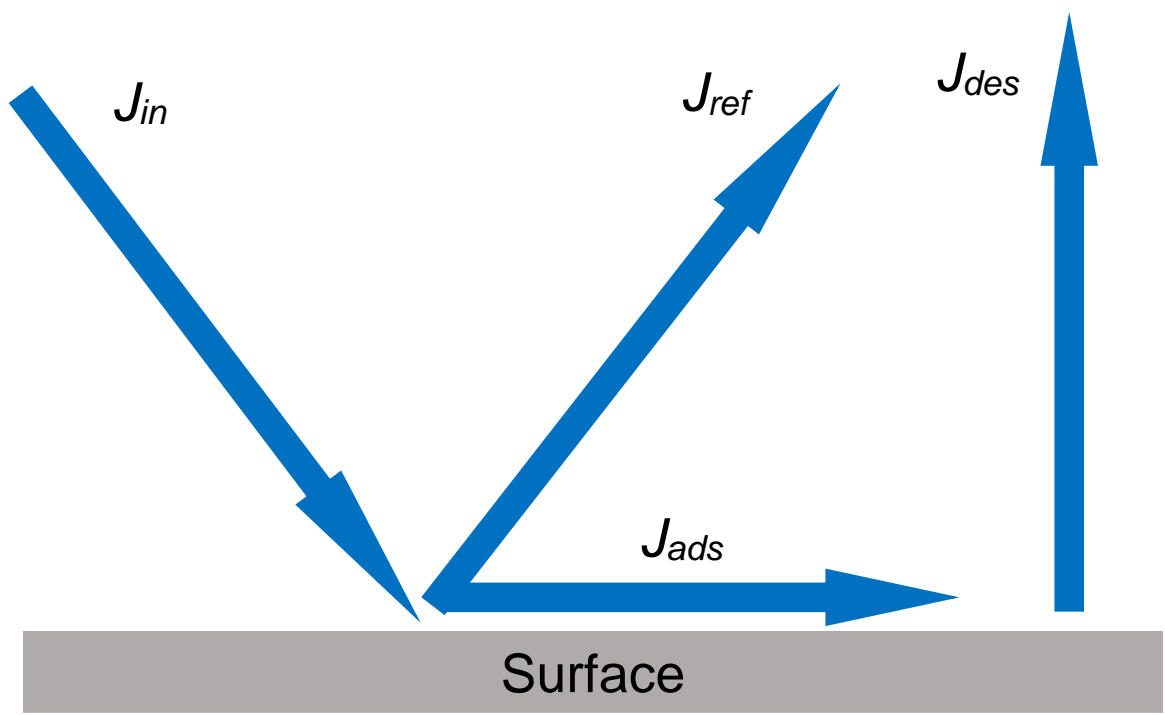

**Figure 6:** Balance of fluxes at a surface; $J_{\text{in}}$ is the input flux, $J_{\text{ref}}$ is the reflected flux, $J_{\text{ads}}$ is the adsorbed flux and $J_{\text{des}}$ is the desorption flux.

Measurements of $S$, or $\alpha$, are fundamental to understanding adsorption rates on surfaces. How such measurements are made depends strongly on the likely magnitude of S and α. The most direct and well-established approach is the King and Wells Method [117]. However, this is limited to measurements of $S$ or $\alpha$ in excess of 0.05 as a consequence of the noise level typically encountered in the partial pressure measurements required to determine $S$ or $\alpha$. Though developed for flat, single crystal surfaces, the King and Wells method has the potential to be used to investigate particulate films deposited on flat surfaces. This is nicely illustrated by measurements by He *et al.* on the sticking of various astrophysically relevant gases on the compact, amorphous solid water surface [118]. The Line-of-Sight Sticking Probability (LoS-SP) Method, developed by Jones and co-workers [119], dispenses with the molecular beam source and provides a somewhat simpler experimental approach to the measurement of $S$ or $\alpha$ of a similar magnitude to that derived from King and Wells measurements.

For smaller values of *S* or *α*, construction of an *adsorption isotherm*, *i.e.* a measurement of surface concentration *versus* the exposure of the surface to the adsorbate gas at fixed temperature is necessary. For flat surfaces, and surfaces with particulate films, this is easily achieved using a calibrated surface spectroscopy (*e.g.*, RAIRS, XPS or even TPD) to probe the surface concentration as a function of the time for which the surface is exposed to the gas at fixed pressure. *S* or *α* (in the absence of desorption) is then defined as

$$S = \frac{\mathrm{d}n_{\mathrm{ads}}}{\mathrm{d}(J_{\mathrm{in}} t_{\mathrm{exp}})} \qquad (6)$$

where $t_{\mathrm{exp}}$ is the time that the surface is exposed to the incident flux. Powdered materials are also open to adsorption isotherm investigations. Indeed, there are well-established gravimetric and volumetric methods for investigating the adsorption of gases on powders that originated in the heterogeneous catalysis community [120,121], which have yet to be applied to powdered synthetic dust grain mimics.

Measurements of adsorption isotherms also open a path to understanding the surface area of the substrate as nicely illustrated by the works of Allouche *et al.* on carbon monoxide (CO) uptake on various amorphous solid water films [122] and of Potapov *et al.* on water ($H_2O$) and CO uptake on carbonaceous films [49,50] using transmission FTIR spectroscopy. The former develops the idea that isotherms derived from studies of gas adsorption equilibria, in particular for multilayer systems, may be interpreted within the Brunauer, Emmett, and Teller (BET) framework [123]. The latter works highlight a real need to understand the surface area of particulate models of dust grains as we may be significantly under-estimating the grain surface area in astrophysical environments. There is, of course, an extensive literature investigating practical adsorbents from which those studying astrophysically relevant materials may learn [124-126].

Measurements of uptake on surfaces whether by King and Wells, LoS-SP or isotherm methods give access to the full variation of *S* or *α* as various binding sites on the surface are filled. This variation can provide a window onto the nature of the adsorption process. Where adsorption proceeds through simple random, empty site filling, as per *Langmuir Adsorption* [127], *S* or *α* is found to depend linearly on the surface concentration of the adsorbate, $n_{\mathrm{ads}}$,

$$S = S_0 \left(1 - \frac{n_{\mathrm{ads}}}{n_{\mathrm{ads,mono}}}\right) \qquad (7)$$

where *S₀* is the initial sticking probability and $n_{\mathrm{ads,mono}}$ is the surface concentration at the saturated monolayer. The ratio $n_{\mathrm{ads}}/n_{\mathrm{ads,mono}}$ is commonly known as the *coverage*, *θ*. *S₀* is a dynamical quantity that may show evidence of adsorbate internal and translational energy and adsorbent surface temperature. Kisliuk developed our understanding of adsorption by recognising that there may be *precursor states* to the final adsorbed state [128,129]. Two types of precursor, *intrinsic* and *extrinsic* precursors, were identified as illustrated in **Figure 7**.

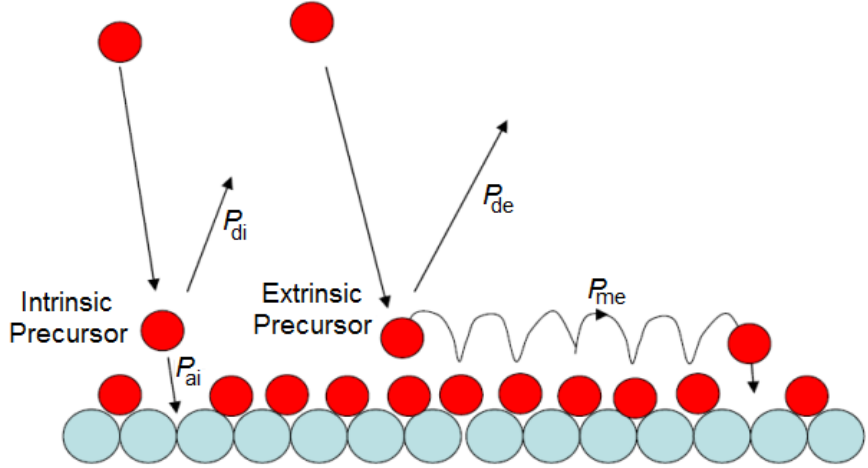

**Figure 7:** Adsorption into the precursor state over empty (*intrinsic*) and filled (*extrinsic*) sites, together with the probability of occurrence of several types of events during its lifetime on the surface. The intrinsic state can adsorb ($P_{ai}$), desorb ($P_{di}$), or diffuse to another site (not shown) from adsorption over an empty site at the surface. On the other hand, the extrinsic precursor adsorbs on parts of the surface with adjacent filled sites. It can then either desorb ($P_{de}$) or diffuse to an adjacent site ($P_{me}$) before eventually adsorbing at an empty site. Reproduced with permission from [129]. © IOP Publishing. All rights reserved.

While Kisliuk [128] considered these precursors as physisorbed states to a final chemisorbed state, it would be appropriate in astrophysical situations to consider the precursors as only partially accommodated species akin to hot atoms/molecules at the surface. Kisliuk was able to show that $S$ or $\alpha$ exhibit the following variation with coverage:

$$S = S_0 \left[ \frac{1 - \frac{n_{ads}}{n_{ads,mono}}}{1 + (K-1)\frac{n_{ads}}{n_{ads,mono}}} \right] \quad (8)$$

where $K$ is essentially an equilibrium constant describing the equilibrium between the gas phase, the precursor state and the adsorbed state.

In addition to revealing details of the adsorption process, adsorption isotherm measurements made as a function of surface temperature can provide a window on the energetics of adsorption [123,130]. Application of van't Hoff analysis to isosteres (lines of constant coverage) derived from a family of isotherms recorded at different temperatures can directly yield the isosteric enthalpy of adsorption, $\Delta_{ads}H$, which can be related to the adsorbate binding energy, $E_b$:

$$\Delta_{ads}H = E_b - RT \tag{9}$$

at fixed temperature for a mole of adsorbate.

Adsorption of hydrogen atoms to carbonaceous and siliceous surfaces is one of the simplest and most studied processes in astrophysically relevant surface science. It is widely accepted that hydrogen ($H_2$) formation in the interstellar medium is mediated, in part, through such surfaces [131,132]. The sticking coefficient, of course, plays a central role in determining the overall rate of this process and hence the concentration of $H_2$ in astrophysical environments [133]. However, interest in such processes on carbonaceous surfaces has blossomed due: (i) to the development of plasma CVD technologies for diamond film growth and (ii) to the application of graphite as a wall lining material in tokomak fusion reactors. With respect to the latter, surface erosion and formation of polycyclic aromatic hydrocarbons acting as a trap for the tritium fuel in such reactors has spurred significant experimental and computational investigation of H interactions with extended π system carbon surfaces at energies above those commonly found in the cold, dense ISM [112,134-138]. These energies explore the chemisorption of H atoms on the unsaturated surfaces and reveal that this might also be an efficient pathway of $H_2$ formation in the warm, diffuse ISM. Studies of radical interactions with carbon surfaces reflect the interest in the latter [139].

### 6.2. Mobility of Molecules on Surfaces

Diffusion of atoms and radicals on surfaces is a fundamental aspect of surface reactivity and has been extensively studied on flat surfaces [140,141]. Langmuir-Hinshelwood chemistry is limited by diffusion rates, while ballistic migration is known to limit Kasemo-Harris reaction rates (see **Section 7** for an introduction to types of reactions at a surface). Classically, we can view diffusion as a simple hoping process with the adsorbate hoping from site to site expressing a random pathway over the surface. This can be represented on a simple, one dimensional diffusion potential, **Figure 8**, as a hop over the barrier presented by the potential from one well to an adjacent well. Such ideas form the basis of classical molecular dynamics [142] and kinetic lattice gas models for processes on surfaces [143] when extended to two dimensions.

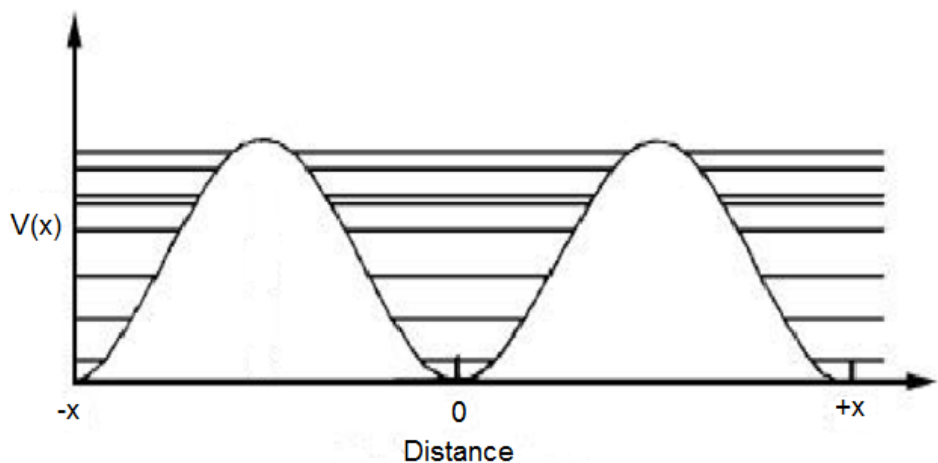

**Figure 8:** A simple one-dimensional periodic diffusion potential showing adsorption sites at 0 and ±x. The horizontal lines represent bound vibrational energy levels within the potential. Such vibrations are essentially frustrated translations. Energy levels exist above the potential barrier but are not shown. These would represent free translation over the surface.

Quantum mechanically such a potential must possess bound vibrational states corresponding to the restricted or frustrated translations and states above the barrier corresponding to free diffusion. The former are represented in **Figure 8** and immediately suggest an alternative to the over the barrier hoping if we accept continuity of the wave functions associated with each energy level. Essentially, this figure posits the idea of resonant tunnelling through the barrier as a means of diffusion. Of course, this is more likely with light atoms but even heavier species can tunnel near the top of the potential. If the diffusion potential can be computationally characterised, it may be possible within this framework to *ab initio* estimate nanoscale diffusion coefficients. The effects of surface morphology (steps, kinks, grain boundaries) might not be so readily treated by such an approach but comparison of computed diffusion coefficients with those measured over microscopic scales may reveal the role of these aspects of diffusion.

Several experimental techniques have been developed to measure diffusion on flat surfaces [144,145]. Such techniques allow measurements on a variety of length and timescales from the looking at individual adsorbate hops on video frame rate timescales at low coverage [146-149] corresponding to diffusion rates of $10^{-19}$ to $10^{-16}$ cm$^2$ s$^{-1}$; through intermediate length scales of tens of nanometre [150,151] to the micrometre scale [152-156] corresponding to diffusion rates from $10^{-15}$ cm$^2$ s$^{-1}$ to about $10^{-5}$ cm$^2$ s$^{-1}$ and providing key information on the energetic barriers for diffusion on larger length scales at a wide range of coverage. Several of these techniques have the potential to be adapted to powdered materials though specific techniques have been developed by the catalysis community to address diffusion on fractal surfaces [157].

Diffusion studies on dust surfaces are largely uncharted territory and present an important future direction as highlighted in **Section 9** of this review.

### 6.3. Desorption Processes from Surfaces
### 6.3.1. Thermal Processes

Within astrophysical environments, several mechanisms can contribute to desorption of species from dust grain surfaces. By far the most studied is thermal desorption as since the 1990s, temperature programmed desorption has become the *de facto* method for experimentally determining adsorbate binding energies, $E_b$, in laboratory astrophysics. Thus, non-equilibrium (kinetic or dynamical) measurements of the rate of desorption, $r_{des}$, given by the Polanyi-Wigner equation [158]:

$$r_{des} = \frac{dn_{ads}}{dt} = \nu(n_{ads})^m e^{-E_{des}(n_{ads})/RT} \qquad (10)$$

where $m$ is the order of desorption, $n_{ads}$ is the surface concentration, and $\nu$ is the surface concentration dependent pre-exponential factor; are alone sufficient to determine $E_{des}$ and its surface concentration dependence. Such dependence reflects the balance of adsorbate-adsorbate and adsorbate-surface interactions.

TPD is a long established technique in surface science [159,160] which has been revolutionised in recent years by the adoption of line-of-sight mass spectrometry methods developed by Jones [119]. Simply, under conditions of high pumping speed, changes in partial pressure measured by quadrupole mass spectrometer can be shown to be proportional to $r_{des}$. While monitoring the appearance of material in the gas phase is the most direct approach, equivalent data can be obtained using measurements of surface concentration loss using infrared spectroscopy or X-ray photoelectron spectroscopy.

Analysis of TPD data is well-described in the literature and given the order of desorption can be approached through a hierarchy of simplifications [161]. These stretch from the simplest assumption of a surface concentration independent $E_{des}$ obtained from simple Arrhenius-like analysis at a fixed coverage through inversion analysis yielding monolayer $E_{des}$ distributions assuming a fixed pre-exponential factor based on Redhead [162] and Hasegawa *et al.* [163] of $10^{12}$ s$^{-1}$ through the recent work of Kay and co-workers [164-166] on the inversion method to Smith *et al.* [167,168] where an optimised pre-exponential factor and monolayer $E_{des}$ distribution are obtained.

The first step in interpreting TPD data is to determine the order of desorption, *i.e.* the dependence of the desorption rate on the surface concentration of the adsorbate. For materials of astrophysical interest, two orders are commonly observed; zeroth order and first order. Zeroth order desorption is observed in desorption of multilayer films. It is characterised by a commonality of leading edges and an increasing peak temperature as increasing amounts of material are deposited on the surface [159]. Desorption from monolayer and sub-monolayer quantities of materials gives first order desorption. This can be characterised by a common peak temperature for desorption with increasing surface concentration when the system exhibits low heterogeneity (*i.e.* a limited range of $E_{\text{des}}$). Where there is significant heterogeneity, common trailing edges and peak temperature decreasing with increasing surface concentration is not uncommon. Similar behaviour is also typical of second order desorption which is commonly associated with recombinative desorption. Though characterisation of the data is the first step in identifying the order of desorption, King [161] describes an application of the initial rate method that allows quantification of order.

Collings *et al.* [169] nicely illustrate the importance of understanding the order of desorption for a number of simple species on amorphous silica grains. **Figure 9** shows the key results from this work; the simple gases, CO, $O_2$ and $N_2$ (**Figures 9(A)** through **(F)**), exhibit common trailing edges and a maximum temperature moving to lower temperature with increasing amounts of material on the silica surface. This is consistent with first order desorption from a growing monolayer and a significant heterogeneity in the binding energy of sites on the surface. Monolayer adsorption is consistent with the adsorbate wetting the surface, *i.e.* the adsorbate is more strongly bound to the substrate than to itself. These materials then with increasing coverage beyond the monolayer show classic multilayer behaviour with common leading edges and a temperature maximum moving to higher values.

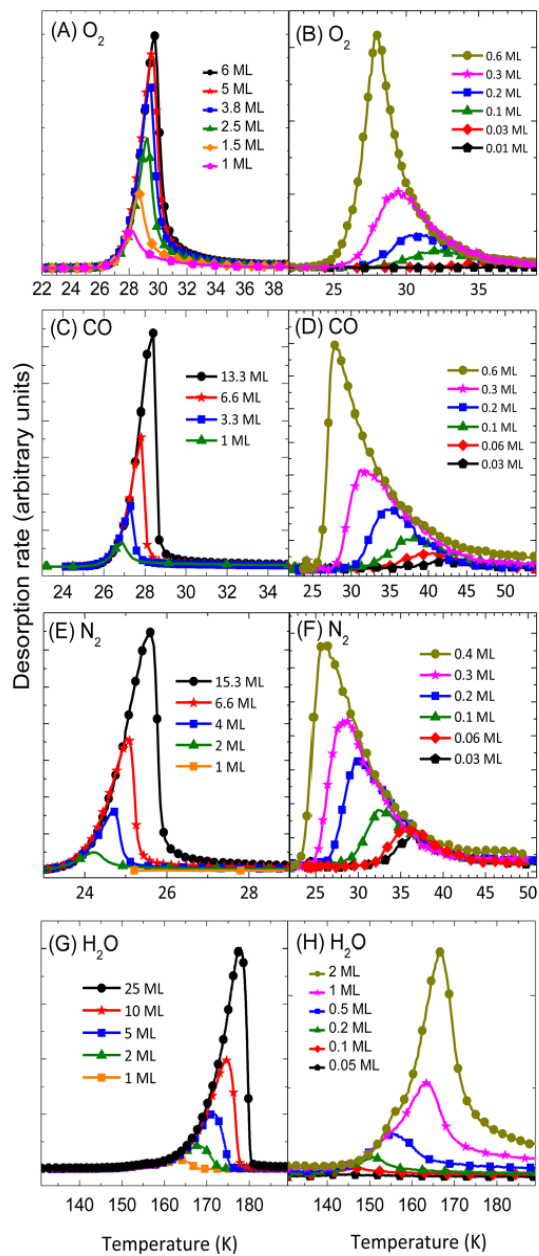

**Figure 9:** TPD profiles of (A-B) $O_2$, (C-D) CO, (E-F) $N_2$ and (G-H) $H_2O$ on amorphous silica for high coverage on the left and low coverage on the right. Reproduced from [169] (Figure 1).

The startling contrast comes with $H_2O$ (**Figures 9(G)** and **(H)**). $H_2O$ shows no change in behaviour with coverage; desorption is consistent with multilayer desorption at all coverages. This suggests that $H_2O$ molecules prefer to interact with other $H_2O$ molecules rather than the silica surface. $H_2O$ does not wet the surface and the silica surface can be considered hydrophobic. Thus, TPD of sub-monolayer quantities of $H_2O$ ice is a good method to investigate the presence and maybe even distribution of hydrophilic binding sites on surfaces (particularly, silicate and carbon surfaces if considering astrophysically relevant materials).

A similarly counter-intuitive result arises in looking at the interaction of benzene ($C_6H_6$) with the amorphous silica surface [101,170,171]. In this case, $C_6H_6$ shows clear monolayer growth and there is even evidence of a second strongly bound layer growing in before the multilayer appears. The $C_6H_6$ is clearly wetting the surface; a result which is consistent with the idea that the surface appears hydrophobic and not our intuitive chemical sense of "like likes like"!

Of course, it is often assumed that $H_2O$ deposition especially on a substrate at temperatures below 20 K is ballistic [172,173], *i.e.* molecules land on the surface and are frozen in place. One might therefore expect some aspect of monolayer growth at low coverage and this raises an immediate question. When does the $H_2O$ become mobile on the surface with the result that we always observe zero order desorption kinetics? Rosu-Finsen *et al.* answered that question in a series of experiments looking at the agglomeration of low doses of $H_2O$ on both amorphous silica and graphite surfaces under ultrahigh vacuum (UHV) conditions [47]. Using reflection-absorption infrared spectroscopy, the authors were able to follow the isothermal temporal evolution of the O-H stretching band as $H_2O$ agglomeration occurred. **Figure 10** illustrates some of their data.

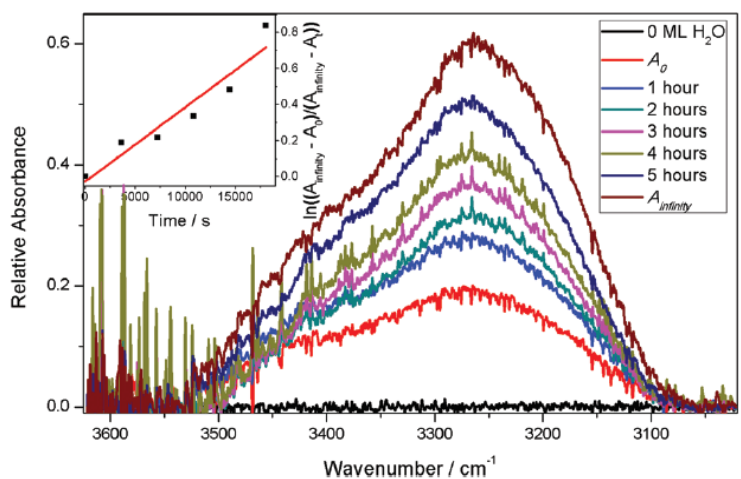

**Figure 10:** Time-resolved RAIR spectra of the O-H stretch region of 0.5 ML equivalent of $H_2O$ on amorphous silica at 18 K. The time between each RAIR spectrum was one hour as this was determined to be the average time between scans. The sharp peaks are due to gas-phase $H_2O$ in the optics boxes on the air side of the UHV apparatus. Inset is a first order kinetic analysis yielding the rate constant for agglomeration at this temperature. Reproduced from [47] with permission from the Royal Society of Chemistry.

Since there is no additional $H_2O$ being deposited during the experiment, the increase in intensity of the O-H stretching band can only be explained in terms of enhanced hydrogen bonding as the $H_2O$ agglomeration takes place and agglomerates grow in size. A simple first order kinetic analysis for each isothermal experiment then allowed construction of the Arrhenius analysis in **Figure 11**. This suggests two regimes of behaviour on the amorphous silica surface; an activated diffusion, with a 2 kJ mol$^{-1}$ barrier, below 25 K and barrierless diffusion above 25 K. Unfortunately, on the HOPG surface, measurements could not be performed below 25 K but the data above 25 K are not inconsistent with an equally barrierless diffusion.

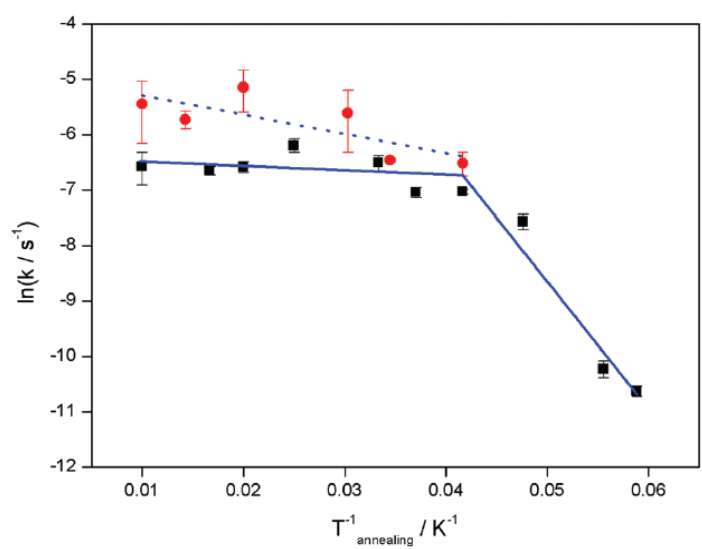

**Figure 11:** Arrhenius analysis of the kinetics of $H_2O$ agglomeration of amorphous silica (filled squares) and HOPG (filled circles) with lines representing the best linear fits for the two surfaces. Reproduced from [47] with permission from the Royal Society of Chemistry.

TPD of $H_2O$ ice mixed with amorphous carbon and silicate grains has been studied by Potapov *et al.* [49]. It was shown that variations of the grain/ice mass ratio lead to a transformation of the TPD curve of the $H_2O$ ice, which can be perfectly fitted with the Polanyi-Wigner equation by using fractional desorption orders. Such an approach has been previously used to describe TPD curves for $H_2O$, $NH_3$, and $CH_3OH$ ices on HOPG [174-176]. The fractional desorption orders of these ices were attributed to hydrogen-bonded networks, which result in a relatively strong interaction between adjacent molecules. In [49], for carbon grains/water ice mixtures, the desorption order of $H_2O$ ice increased from 0 for pure $H_2O$ ice to 1 with the increase of the grain/ice mass ratio. For two silicate grains/water ice mixtures studied, the desorption order of $H_2O$ ice was obtained to be 1. These results were explained by desorption of $H_2O$ molecules from the large surface area of highly-porous aggregates composed of carbon or silicate grains (see **Figure 4** of this review). Further study, on TPD of CO ice mixed with amorphous carbon grains [50] confirmed the desorption of ice molecules from the extensive surface of the grains and allowed the authors to make a conclusion about the importance of dust surfaces in physico-chemical processes in astrophysical environments. In addition, in [49] it was shown that there is a distribution of binding energies of $H_2O$ molecules on silicate grains providing evidence of hydrophilic and hydrophobic surface binding sites.

Kinetic analysis of zeroth order data is relatively simple. The surface concentration term disappears from equation (10) and it is possible to obtain $E_{des}$ directly from an Arrhenius plot of $\ln(r_{des})$ *versus* 1/T. For first order desorption, a starting point is often the work of Redhead [162] which allows a single $E_{des}$ to be derived assuming $\nu_{des}$ of $10^{12}$ s$^{-1}$ for physisorption. However, it is becoming increasingly clear that the assumption of a single value for $E_{des}$ is no longer valid, and we must apply direct inversion of the Polanyi-Wigner equation (10). This gives $E_{des}$ as function of the surface concentration at time *t*, *N(t)*:

$$E_{\text{des}} = -RT \ln\left(\frac{\mathrm{d}n_{\text{ads}}/\mathrm{d}t}{\nu n_{\text{ads}}(t)^m}\right) \tag{11}$$

To determine $n_{ads}(t)$, the initial surface concentration ($n_{ads,tot}$) is assumed to be given by the rate of bombardment ($Z_W$) multiplied by the dose time ($\tau$) (equation (12)):

$$n_{\text{ads,tot}} = Z_w \tau = \frac{PS\tau}{\sqrt{2\pi m k_B T}} \tag{12}$$

where *P* is the pressure, *S* is the sticking coefficient and *m* is the mass of the adsorbate. At low surface temperatures, the flux of species leaving the surface either directly upon collision or desorbing is believed to be relatively small, making *S* near unity. The values of $n_{ads}(t)$ are obtained by subtracting the total gas phase concentration at the previous time step from the initial surface concentration ($n_{ads,tot}$). The values of $\mathrm{d}n_{ads}/\mathrm{d}t$ are determined by the experimental TPD data. Within the sub-monolayer regime, *m* is assumed to be 1. Within the original inversion method as described by Tait *et al.*, a fixed value of $\nu_{des}$ of either $10^{12}$ or $10^{13}$ s$^{-1}$ is assumed [164-166] and plots of $E_{des}$ against $n_{ads}(t)$ are constructed for each adsorbate sub-monolayer dose on the substrate and averaged to give the full range of $E_{des}$ *versus* $n_{ads}$ on the substrate. A functional fit is then made to the averaged $E_{des}$ *versus* $n_{ads}$ data to obtain the $E_{des}(n_{ads})$ function. TPD simulations are made using this function and simply compared to the experimental data.

The extended inversion method of Smith *et al.* [167,168] departs from this formulation by using a fixed, but variable, value of $\nu_{des}$ in the range $10^{12\pm10}$ s$^{-1}$. Again, plots of $E_{des}$ against $n_{ads}(t)$ are constructed for each adsorbate sub-monolayer dose on the substrate and averaged to give the full range of $E_{des}$ *versus* $n_{ads}$ on the substrate. The TPD is simulated and residuals estimated such that the optimised $\nu_{des}$ is obtained through minimisation of the residuals. Finally, the optimised $\nu_{des}$ would be used to derive the optimised TPD, which would be modelled using this optimised $E_{des}(n_{ads})$ function for the system and to provide a final comparison of simulated TPD with experimental data.

The inversion technique of Kay and co-workers [164-166] has been widely adapted to describe small molecule desorption from many heterogeneous astrophysically-relevant surfaces including carbonaceous [174-189], siliceous [101,169-171,181,190-195] and ice surfaces [196,197]. But we are now beginning to see the extended inversion method of Smith *et al.* coming to the fore with studies likewise on a variety of carbonaceous [168], siliceous [167,198,199] and ice surfaces [168,198,200]. As an illustrative example, we consider the study of CO desorption from amorphous silica by Taj and co-workers [198,199]. **Figure 12** shows the $R^2$ (the sum of the squared error of the fit) data *versus* pre-exponential factor for these measurements. This yields an optimised pre-exponential factor that is used to recover the $E_{des}$ *versus* coverage curve shown in **Figure 13**. In this particular study, the coverage dependence of $E_{des}$ and its probability distribution, $P(E_{des})$,

$$P(E_{des}) = -\frac{dn_{ads}}{dE_{des}} \qquad (13)$$

were then used to describe the heterogeneity of the surface in undertaking CO IR line-profile synthesis at 18 K. This in turn revealed that the line-profile could only be simulated assuming an adsorb and diffuse mechanism for filling the monolayer, *i.e.* that the CO molecules explore the heterogeneous surface filling that surface from the most strongly bound sites to the least strongly bound as the coverage increases. It also revealed that the line width of the CO vibrational band, 2.3 cm$^{-1}$, suggests a vibrational lifetime of 1.15 ps on the amorphous silica surface perhaps consistent with vibrational predissociation of the CO from the surface.

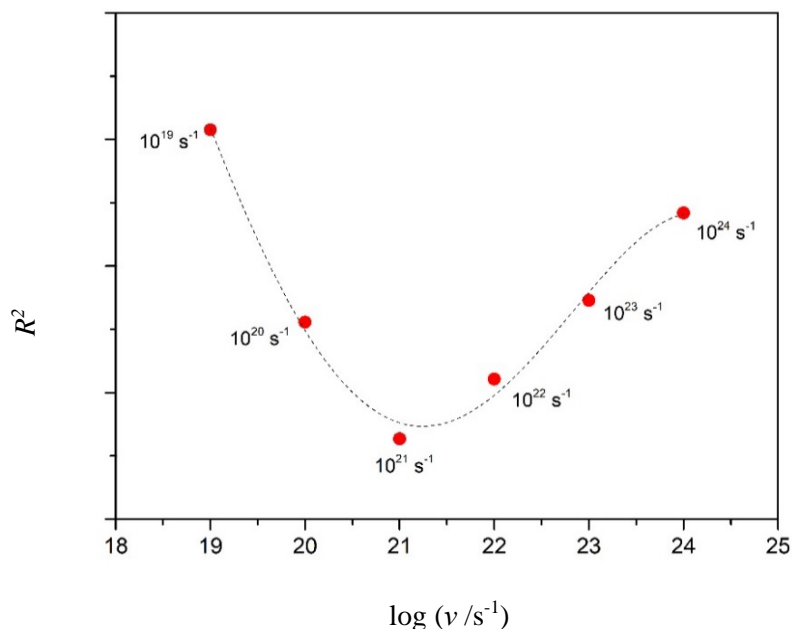

**Figure 12:** A plot for $R^2$ derived from experimental and simulated TPD of CO from amorphous silica *versus* the log of the pre-exponential factor used in the inversion analysis. This yields, at the minimum of $R^2$, a corresponding to a value of $\nu$ of $1.74^{+0.75}_{-0.53} \times 10^{21}$ s$^{-1}$. Reproduced from [199] (Figure ESI-1).

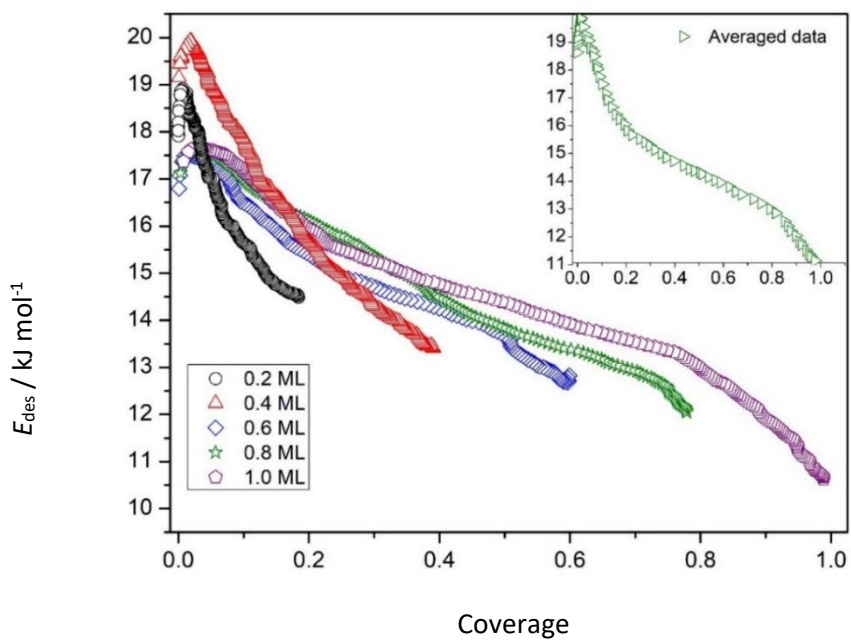

**Figure 13**: $E_{\text{des}}$ *versus* coverage for sub-monolayers of CO desorbing from an amorphous silica substrate derived using $\nu$ of $1.74^{+0.75}_{-0.53} \times 10^{21}$ s$^{-1}$. The inset presents the averaged data over all coverages. Reproduced with permission from [199] (Figure 1a).

This extensive range of work utilising inversion methods emphasises that the common assumption that desorption from a complex substrate can be described by a single activation energy for desorption is simply not true. Both heterogeneity in the adsorbate-substrate interaction, reflecting the variety of binding sites on the surface, and adsorbate-adsorbate interactions, reflecting the coverage dependence of the adsorption process, play a role in determining the activation energy for desorption.

While TPD dominates the investigation of thermal desorption from surfaces, isothermal desorption is sometimes utilised. In this approach, the adsorbate loaded surface is rapidly heated to a fixed, but variable, temperature where desorption is known to occur. The change in the gas phase pressure with time [201] or of the surface concentration of the adsorbate with time [202] is measured to yield a desorption rate. The surface temperature dependence of that rate can then be determined and hence $E_{des}$ derived.

The availability of good quality $E_{des}$ data from experimental measurements makes it possible to benchmark computational approaches to investigating adsorption, desorption and reaction on model surfaces. We do not propose an exhaustive review of this computational literature. Rather we present some selective examples to illustrate both methodologies and their capabilities on some simple systems. Three common computational approaches have been applied to the astrophysically relevant problems; bare cluster approaches, embedded clusters and periodic method. All of these typically utilise density functional theory (DFT) methods.

The bare cluster approach is nicely illustrated through studies of processes simulating the graphene and graphite surface using the polycyclic aromatic hydrocarbon coronene as a substrate for the reaction [203,204]. Here, it is possible to completely describe the system quantum mechanically and derive adsorption energetics and reaction barriers. Application of transition state theory then gives the relevant rate constants.

Embedding the quantum mechanical cluster in an extended cluster described using the molecular mechanics approach takes us to the QM/MM approach. This has the added advantage over the simple cluster method that energy dissipation within a much larger volume is possible bringing the system more closely to an extended surface system. This approach is particularly useful for studies on three dimensional crystalline and amorphous materials where the QM region would include the adsorbate(s) and silica [205-207] or silicate [208,209] unit which is embedded in an extended silica/silicate MM framework.

Finally, as we have seen an evolution in computer resources, the application of periodic DFT methods to explore the energetics of adsorption, reaction and desorption processes has come to the fore [210-213]. Such simulations bring the tools of solid state physics to the table and potentially open new ways of understanding the astrochemistry of surfaces. However, the simulation of amorphous materials using periodic DFT is challenging in requiring very large amorphous unit cells to describe the extended amorphous materials.

As a final comment on thermal desorption, it should be recognised that there an inherent limitation with these techniques in relation to the measurement of atom and radical $E_{des}$. We can trap such species on cryogenic (model dust grain) surfaces but the heating of that surface to drive thermal desorption enables surface mobility and will promote reaction. Any thermal desorption measurement is therefore likely only to see the products of those reactions. One experimental approach is open to us in that case and that is collision-induced desorption (CID) [214]. The impact of a fast, rare gas atom with a surface can, through collisions with adsorbates on the surface, promote desorption of those adsorbates. Measurements of adsorbate surface concentration as a function of time can then yield the CID cross-section. In turn, measurements of this cross-section as a function of the impact energy of the rare gas will yield a threshold from which $E_{des}$ can be determined [214]. These challenging measurements potentially provide a route to $E_{des}$ for atoms and radicals which could be used to benchmark the most sophisticated computational studies on such systems.

### 6.3.2. Non-thermal Processes

While there is a substantial literature relating to thermal desorption from model dust grain surfaces, the literature relating to non-thermal processes is more limited and for the most part has focussed on the role of non-thermal desorption from molecular solid surfaces [215,216]. Most non-thermal processing of grain surfaces has focussed on the chemical transformations that occur in mixtures of simple ices and the rise of chemical complexity [217].

Surface photochemistry and charged particle induced chemistry are largely focussed on metallic surfaces. In that regime, the role of excitation of electron-hole pairs both below and above the work function of the metal is central in desorption induced by electronic transition (DIET) processes. There are numerous reviews of this literature which has blossomed in recent years with the development of ultrafast laser sources permitting real-time investigation of the fate of the surface electronic excitation [218-224]. The behaviour of graphitic surfaces is probably not dissimilar to that of metallic surfaces. While some carbonaceous and siliceous materials, as insulators, will have a significantly simpler non-thermal chemistry as there are no routes to excitation of metal-like electron-hole pairs in these systems. An illustrative example of this difference arises when the photochemistry of water on graphite [225,226] is compared with that on amorphous carbon foils [227]. In the former, hot electron processes result in water dissociation and etching of the graphite surface to carbon monoxide. On the latter, simple desorption is reported.

A few examples on UV photodesorption of ice molecules from the surface of cosmic dust analogues can be given. The photodesorption yield of $H_2O$ molecules was measured for ice coatings on amorphous carbon foils [227] and amorphous carbon and silicate grains [228]. From the results presented there, one can conclude that $H_2O$ photodesorbs more efficiently from dust surfaces compared to standard substrates. Very recently, results on photodesorption of $H_2O$ ice mixed with carbon or silicate grains were obtained by Potapov *et al.* [229]. It was shown that mixing water ice with dust grains enhances photodesorption of water compared to the pure water ice at 10, 100, and 150 K. These results have an important consequence in the calculations of the lifetime of icy grains in different astrophysical environments, particularly in protoplanetary and debris disks.

## 7. Chemistry on Dust Surfaces

As with most chemical reactions, we can through careful study establish reaction mechanisms for processes occurring on solid surfaces. In developing such mechanisms, we find that there are three common types of reactions at a surface: Eley-Rideal reactions, Langmuir-Hinshelwood reactions, and Kasemo-Harris reactions.

The first mechanism of these was developed in the 1940s by Eley and Rideal from measurements of *ortho-para* exchange in $H_2$ and isotope exchange between $H_2$ and $D_2$ over tungsten surfaces. The Eley-Rideal reaction can be described in terms of a simple sequence of steps which involve the reaction of an adsorbed species with gaseous reactant as illustrated in **Figure 14**.

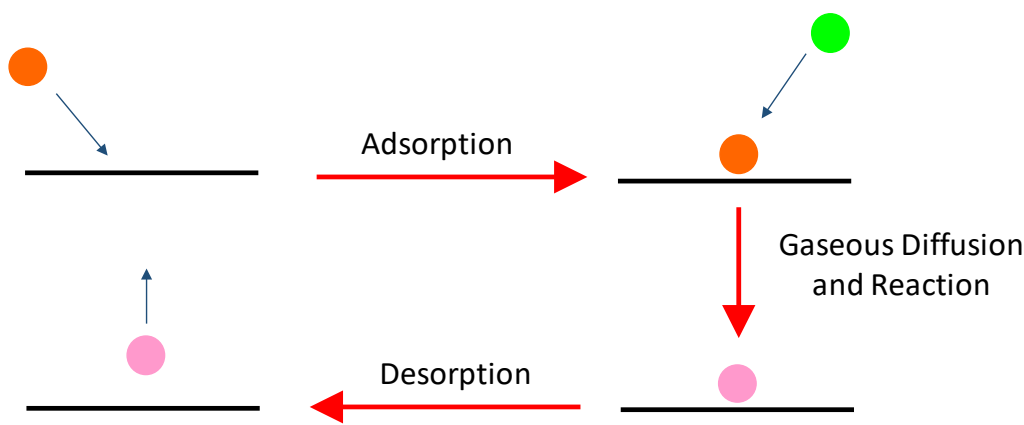

**Figure 14.** The Eley-Rideal reaction mechanism.

In terms of a chemical equation, this mechanism can be written as:

$$A(ads) + B(g) \rightarrow \text{Products} \tag{14}$$

from which it is clear that the rate of reaction can be written as:

$$v_r = k_{ER}\theta_A P_B \tag{15}$$

where $\theta_A$ is the surface concentration of species A and $P_B$ is the partial pressure of species B. Hence applying the Langmuir Isotherm to the adsorption of the reactant A, we have:

$$v_r = k_{ER}\frac{b_A P_A P_B}{1 + b_A P_A} \tag{16}$$

where $b$ is the Langmuir coefficient for A.

There are two limiting situations with this mechanism. At low pressures, and hence low surface concentrations, perhaps consistent with chemistry in astrophysical environments, the general Eley-Rideal rate law reduces to:

$$v_r = k_{ER} b_A P_A P_B \tag{17}$$

and the reaction is first order in both gaseous reactant pressures. At high pressures, and hence high surface concentrations:

$$v_r = k_{ER} P_B \tag{18}$$

and the reaction is zero order in the adsorbed species but first order in the gaseous reactant pressures.

The second mechanism that we consider bears the names of Langmuir and Hinshelwood and involves the reaction of one or two adsorbed species as illustrated in **Figure 15**.

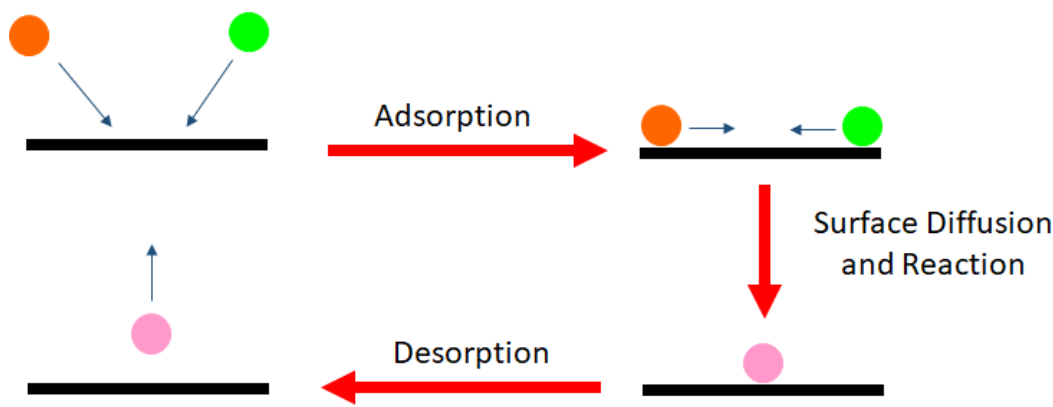

**Figure 15.** The Langmuir-Hinshelwood reaction mechanism.

The first example of such a process is the unimolecular decomposition of an adsorbate which in terms of a chemical equation can be written as:

$$A(ads) \rightarrow Products \tag{19}$$

from which it is clear that the rate of reaction can be written as:

$$v_r = k_{ULH}\theta_A \tag{20}$$

Hence applying the Langmuir Isotherm to the adsorption of the reactant A, we have:

$$v_r = k_{ULH}\frac{b_A P_A}{1 + b_A P_A} \tag{21}$$

There are two limiting situations with this mechanism. At low pressures, and hence low surface concentrations reflecting those found in astrophysical environments, the Unimolecular Langmuir-Hinshelwood rate law reduces to:

$$v_r = k_{ULH} b_A P_A \tag{22}$$

and the reaction is first order in the gaseous reactant pressure. At high pressures, and hence high surface concentration:

$$v_r = k_{ULH} \tag{23}$$

and the reaction is zero order in the adsorbed species.

The second example of such a process is the bimolecular surface reaction of two adsorbed species which in terms of a chemical equation can be written as:

$$A(ads) + B(ads) \rightarrow Products \tag{24}$$

from which it is clear that the rate of reaction can be written as:

$$v_r = k_{BLH}\theta_A \theta_B \tag{25}$$

Hence, applying competitive Langmuir adsorption to the reactants A and B, we have:

$$v_r = k_{BLH}\frac{b_A P_A b_B P_B}{(1 + b_A P_A + b_B P_B)^2} \tag{26}$$

This rate law shows some interesting features. If the interaction of reactant A and reactant B with the surface are comparable in strength, then neither will have overwhelming importance in the denominator and the rate of reaction will go through a maximum as $P_B$ is increased for constant $P_A$, and *vice versa.* At low surface concentrations of A and B (either low gas pressures or weak adsorption and consistent with astrophysical environments), the Bimolecular Langmuir-Hinshelwood rate law reduces to:

$$v_r = k_{BLH} b_A P_A b_B P_B \quad (27)$$

and the reaction is first order in both of the gaseous reactant partial pressures.

At low surface concentrations of A with respect to B (either low gas pressures or weak adsorption of A with respect to B), the Bimolecular Langmuir-Hinshelwood rate law reduces to:

$$v_r = k_{BLH} \frac{b_A P_A b_B P_B}{(1 + b_B P_B)^2} \quad (28)$$

and the reaction is first order in the partial pressure of A and goes through a maximum rate as the pressure of B is increased at constant $P_A$. A similar relationship is derived if the surface concentration of B is small with respect to A.

Our final example of a surface reaction mechanism is the relatively recently identified Hot Atom reaction, which now bears the names of Kasemo and Harris, as illustrated in **Figure 16**.

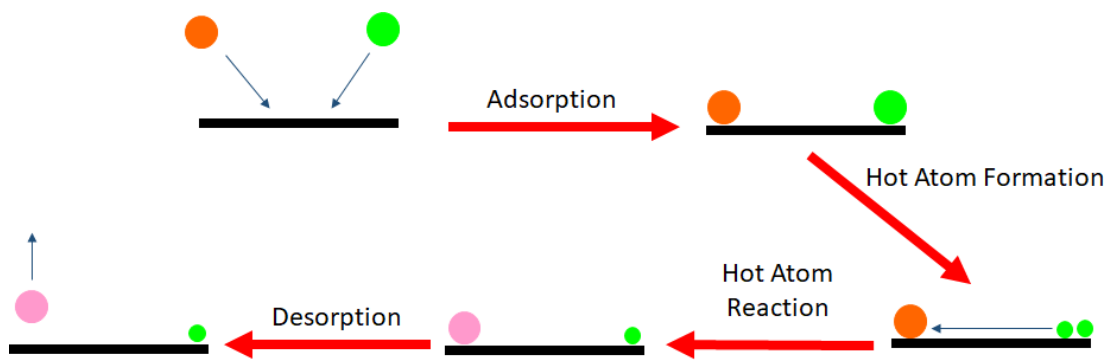

**Figure 16.** The Kasemo-Harris reaction mechanism.

The hot atoms themselves have two potential sources; photodissociation of adsorbed species, *e.g.* $O_2$ (indeed it was *via* studies of $O_2$ photochemical surface reactions that this mechanism was identified) and incomplete accommodation of a hot atom or radical incident from the gas phase on the surface.

The balance of adsorption and desorption moderates the surface concentrations of reactants while chemical species, including atoms such as H, O, and N, can move across the surface even at low temperatures, meet each other and react. Thus, adsorption, desorption, and diffusion processes described in the previous section play an important role in grain surface chemistry.

The main triggers of dust grain surface chemistry in cosmic environments, as illustrated in **Figure 3,** are photon irradiation, cosmic rays interactions (both processes dissociate molecules producing reactive radicals), thermal processing (allowing to reaction barriers to be surmounted), and atom addition from the gas phase. In addition to providing a meeting place for reactant, dust grains show catalytic effects which include: (i) dissipation of energy excess released in bond formation (*i.e.* acting as a third body); (ii) lowering of diffusion barriers for reactants and activation barriers of reactions; and (iii) direct participation (by atoms or functional groups) in surface reactions. In the following, experimental evidence for catalytic effects (i) and (ii) (**Section 7.1**) and (iii) (**Section 7.2**) are presented.

### 7.1. Catalytic Formation of Molecules on Dust Surfaces

Molecular hydrogen ($H_2$) is the most abundant molecule in the Universe and water ($H_2O$) is one of the main species in astrophysical environments (and the main constituent of ices in the ISM, protostellar envelopes and planet-forming disks beyond the snowline). These molecules strongly influence the astrophysical conditions of the environment and are starting points in many key astrochemical reactions.

It is well known that $H_2$ cannot be efficiently formed in the gas phase of the ISM due to the low efficiency of three-body collisions and forbidden ro-vibrational transitions involved in the radiative association of two hydrogen atoms. $H_2$ formation mainly occurs on surfaces of interstellar dust grains. This is true too for water in low-temperature astrophysical environments. While there are gas-phase routes, *via* ion-molecule chemistry, to its formation, the dominant route is by atom recombination on the surface of dust grains. In both cases, the catalytic role of the dust surface is mainly to enable excess energy dissipation and to stabilise the reaction product.

For a review of $H_2$ formation, we recommend [131]. Various studies of $H_2$ [230-238] and $H_2O$ [239-241] formation on silicate and carbon surfaces by H/O atom recombination have illustrated that the reaction efficiency is highlight dependent on surface morphology.

It was found that compared to crystalline silicate grains, amorphous silicate grains are efficient catalysts of $H_2$ formation at temperatures relevant to interstellar clouds [233]. On the basis of analysis of TPD curves, the authors demonstrated a 1.4 – 1.5 times increase in the activation energy for $H_2$ desorption on the amorphous silicate compared with polycrystalline silicate surface. Consequently, the temperature window in which $H_2$ formation is efficient is broadened and shifted by a similar factor. **Figure 17** shows calculated recombination efficiencies of $H_2$ at steady state on amorphous and polycrystalline silicates. A window of high-recombination efficiency was found between 6 and 10 K for polycrystalline silicate and between 9 and 13 K for amorphous silicate. This result has an important consequence in estimating formation yields of $H_2$ in the ISM, where the ratio between amorphous and crystalline silicates is still an open question.

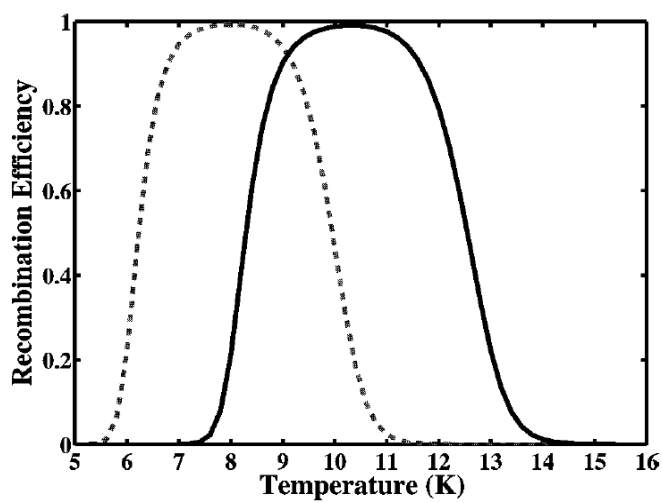

**Figure 17**. Calculated recombination efficiency of $H_2$ at steady state on amorphous silicate (solid line) and polycrystalline silicate (dashed line) *versus* temperature. Reproduced with permission from [233]. © AAS.

Dulieu and co-workers [241] have investigated the formation of $D_2O$ through the reaction of D atoms with adsorbed $O_2$ molecules on amorphous silicate and graphite substrates. The deuterated chemistry was chosen to avoid issues with the $H_2$ and $H_2O$ contamination found in most stainless steel ultrahigh vacuum systems. The work demonstrated that a fraction of the newly formed species were desorbed into the gas phase during their formation due to reaction exothermicity and that this chemical desorption process is sensitive to the type of surface upon which the formation reaction occurs. Chemical desorption of $D_2O$ is evident in this case due to the exothermicity (5.2 eV) of the OD + D reaction in the reaction chain:

$$O_2 + D \rightarrow DO_2 + D \rightarrow OD + OD, \; OD + D \rightarrow D_2O \tag{29}$$

These results highlight the role of chemical desorption in chemistry on dust surfaces and on the abundance of gas-phase and solid-state species in cold astrophysical environments.

In contrast, studies of the reaction of D with adsorbed ozone ($O_3$) on amorphous silicates by He and Vidali [240] presented no evidence for chemical desorption. In this case, the reaction route to water starts from the $O_3$ hydrogenation reaction:

$$O_3 + D \rightarrow OD + O_2 \qquad (30)$$

The analysis in [240] also suggested that OD was likely to co-desorb with $D_2O$ in the TPD experiment. Based on this finding, the authors concluded that the temperature range for the efficient formation of water on amorphous silicate grains should be extended from about 30 K to at least 50 K, and possibly over 100 K, due to a higher binding energy of OH radicals compared to the value typically used in chemical models. This result speaks for the possible efficient formation of water on dust grains in warmer environments, such as protostellar envelopes and planet-forming disks.

Evidently, it is important to understand if the surface properties of silicates (particularly their hydrophobic and hydrophilic properties) are responsible for the existence or non-existence of chemical desorption.

Considering more complex species, Fischer–Tropsch synthesis (indirect catalytic hydrogenation of the oxides of carbon) of methane ($CH_4$) and $H_2O$ from CO and $H_2$:

$$CO + 3H_2 \rightarrow CH_4 + H_2O \qquad (31)$$

and Haber–Bosch synthesis of ammonia ($NH_3$) from $N_2$ and $H_2$ (normally achieved at very high pressures, 70 – 350 times atmospheric pressure, and temperatures of a few hundred ºC in the presence of a metal catalyst):

$$N_2 + 3H_2 \rightarrow 2NH_3 \qquad (32)$$

have been studied on amorphous silicate surfaces at high temperatures, 500 – 900 K, using a combination of gas phase Fourier transform infrared (FTIR) spectroscopy and gas chromatography mass spectrometry (GC-MS) to monitor the reaction products [242]. Strong evidence of significant catalytic efficiency of silicates was revealed. In the case of experiments involving CO, $N_2$, and $H_2$, reactions on the silicate surface led to the formation of COMs of prebiotic interest, such as methylamine ($CH_3NH_2$), acetonitrile ($CH_3CN$), and N-methyl methylene imine ($H_3CNCH_2$). These results were discussed as being relevant to the formation of molecules including COMs in the early Solar Nebula.

Formamide has been detected in molecular clouds and protostellar envelopes [243-245] and in comets [246-248]. The effect of the substrate in the thermal synthesis of purines and pyrimidines from formamide triggered by heat (160º C) on olivine surfaces with different Mg/Fe ratios (from fayalite to forsterite) was studied using GC-MS [249]. These experiments showed that olivines favour formamide condensation into various pyrimidine derivatives including two nucleobases providing an example of possible extraterrestrial syntheses of these prebiotic molecules. A condensation scheme of formamide in the presence of cosmic dust analogues is shown in **Figure 18**.

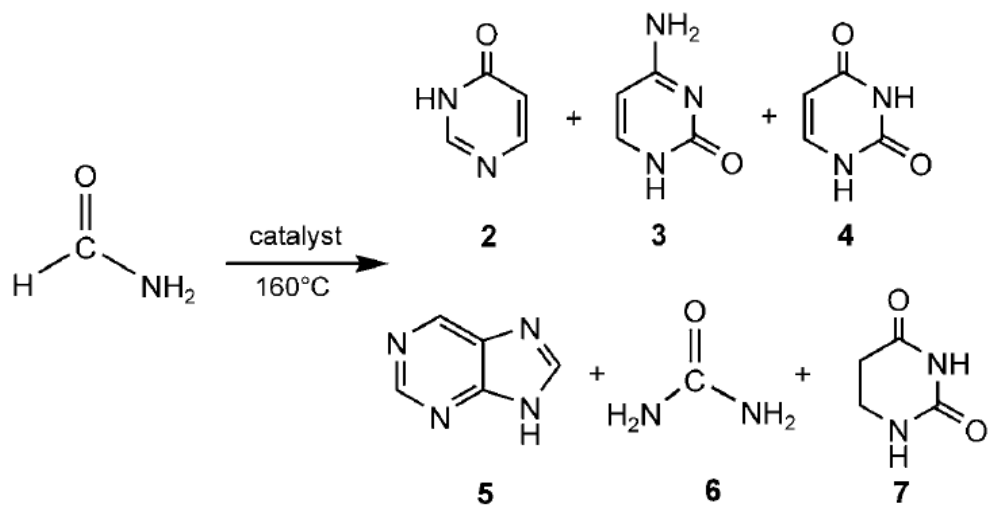

**Figure 18.** Condensation scheme of formamide in the presence of cosmic dust analogues. Products: 2 – pyrimidinone, 3 – uracil, 4 – cytosine, 5 – purine, 6 – urea, 7 – dihydrouracil. Reproduced with permission from [249]. Copyright Wiley-VCH GmbH.

These studies [242,249] are relevant to the chemistry taking place not only in warmer environments, such as inner regions of protostellar envelopes and planet-forming disks but also in the high-temperature atmospheres of exoplanets.

In [250], the abundances of new molecular species, characterised by FTIR spectroscopy, formed from formamide irradiated with 200 keV protons at low temperatures were compared for amorphous olivine (MgFeSiO$_4$) grains and inert silicon surfaces. The comparison showed that the presence of dust grains reduces the overall yield of synthesised species (NH$_3$; CN$^-$; NH$_4^+$OCN$^-$, CO$_2$, HNCO, CO). This result was linked to the larger surface area of silicate grains compared to that of the silicon substrate, which may enhance desorption of formamide. A similar conclusion was recently reached in a study on photochemistry in H$_2$O:NH$_3$:CO$_2$ ices at 75 – 150 K [251], where it was shown that the formation of COMs on the amorphous enstatite (MgSiO$_3$) grains is decreased compared to a KBr substrate possibly due to the different desorption rates of the reactants from larger area silicate and smaller area KBr surfaces. In addition, in [250], the charge exchange due to presence of iron and magnesium metals in silicates was proposed as one more possible effect.

The next study demonstrating the catalytic effect of the dust surface as compared to an astrophysically non-relevant surface and also showing the catalytic effect of the dust surface in the formation of COMs at low temperatures was published recently by Potapov *et al.* [252]. The thermal reaction:

$$CO_2 + 2NH_3 \rightarrow NH_4^+NH_2COO^- \tag{33}$$

leading to the formation of ammonium carbamate was studied. Ammonium carbamate is a possible precursor of urea. Urea in turn can be a precursor of pyrimidine which is required for the synthesis of nucleobases in RNA molecules [253]. Thus, formation of ammonium carbamate on the surface of cosmic grains in molecular clouds, protostellar envelopes or protoplanetary disks could trigger a network of prebiotic reactions. **Figure 19** shows an example of the IR spectra, taken after the deposition of a $CO_2:NH_3$ mixture at 15 K and after 4 hours of the isothermal kinetics experiment at 80 K, where characteristic IR bands of ammonium carbamate are visible. It was demonstrated that surface catalysis on silicate and carbon grains accelerates the kinetics of the reaction at a temperature of 80 K by a factor of up to 3 compared to the reaction occurring on a standard substrate such as KBr. Reaction rate coefficients obtained from isothermal measurements following the formation of ammonium carbamate for different surfaces and nominal ice thicknesses (calculated for the KBr surface area) are presented in **Figure 20**.

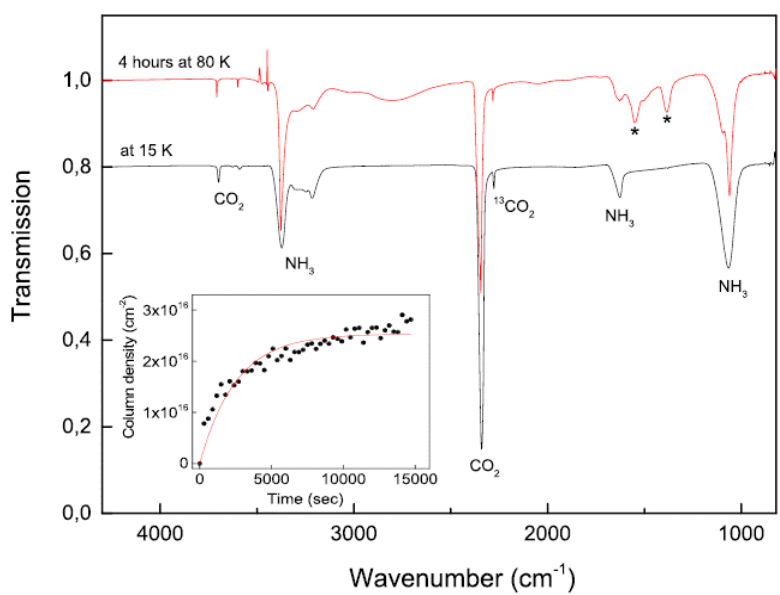

**Figure 19.** IR spectra taken after the deposition of a $NH_3:CO_2$ 4:1 mixture on carbon grains at 15 K and after 4 hours of the reaction at 80 K. Two bands related to $NH_4^+NH_2COO^-$ are marked by asterisks. The 15 K spectrum is vertically shifted for clarity. Inset: The time dependence of the $NH_4^+NH_2COO^-$ column density calculated from the 1550 cm$^{-1}$ band derived from the isothermal kinetic experiments performed with $NH_3:CO_2$ 4:1 ices on carbon grains at 80 K. Reproduced with permission from [252]. © AAS.

With the decrease of the ice thickness, starting from 200 nm nominal thickness, a dramatic increase of the reaction rate coefficient can be observed for carbon and silicate grains pointing to the participation of the dust surface in the reaction studied. Note that around this nominal thickness the monolayer–multilayer transition in $CO_2+NH_3$ ices may take place on the grains due to their high porosity and corresponding large surface area (see [50,252] and **Figure 4** for an illustration). The increased reaction rates for grains compared to those measured for $NH_3:CO_2$ ices on KBr can be explained in two ways. The first explanation is purely physical: at low temperatures the reaction is diffusion-limited and increased diffusion rates of the reactants may lead to the increased reaction rate. The second explanation implies either the formation of an intermediate weakly bound $CO_2$–$NH_3$ complex that helps to overcome the reaction barrier (however, there was no evidence of the formation of the complex in the IR spectra) or a reduction of the reaction barrier. Very recently, the temperature range of the experiments was extended down to 50 K and the catalytic role of the dust grain surface in the $CO_2 + 2NH_3 \rightarrow NH_4^+NH_2COO^-$ reaction was related to a reduction of the reaction barrier [254].

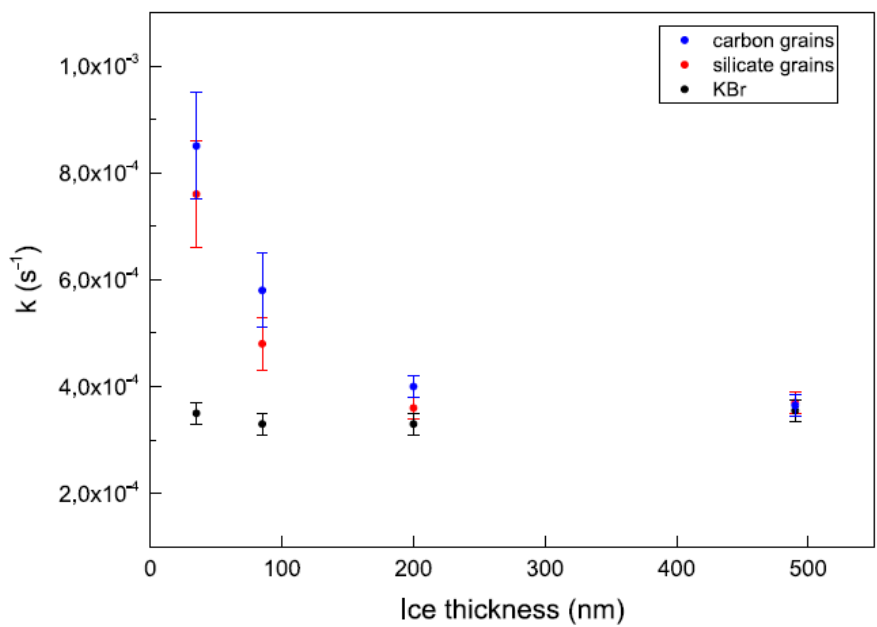

**Figure 20.** Dependence of the pseudo first-order reaction rate coefficients of the $CO_2 + 2NH_3 \rightarrow NH_4^+NH_2COO^-$ reaction on the ice thickness for various surfaces. Reproduced with permission from [252]. © AAS.

However, the authors of [254] could not directly distinguish between reaction and diffusion processes from their measurements. The timescales for the diffusion and reaction were compared meaning that the slower process was measured. The timescale for diffusion was estimated using the measured diffusion coefficients for $CO_2$ and $NH_3$ along pores of amorphous water ice as being much shorter than the experimental timescale. However, the efficiency of diffusion of molecules on porous dust grains having a large surface area can be quite different as compared to standard substrates or water matrixes. This stresses the need for direct measurements of diffusion coefficients for astronomically relevant molecules on reliable cosmic dust grain analogues. Only having such values, we will be able to have a complete picture of the physico-chemical processes on dust surfaces.

### 7.2. Formation of Molecules from Grain Surface Atoms

We start our discussion in this section from the formation of CO and $CO_2$ in $H_2O$ and $O_2$ ices covering carbon grains and foils by photon and ion irradiation. CO and $CO_2$ are the most abundant solid-state species, after $H_2O$, in cold astrophysical environments.

Mennella *et al.* irradiated $H_2O$ ices on hydrogenated amorphous carbon grains by 30 keV $He^+$ ions [255] and Lyman-$\alpha$ (10.2 eV) photons [256] at 12 K. The authors estimated from the experimental results that from a few to tens of percent of the carbon of carbonaceous particles can be converted to CO and $CO_2$ during a dense cloud timescale by ions and UV photons. This demonstrated an additional mechanism (to UV sputtering) for eroding carbonaceous dust grains in cold astrophysical environments – through the formation of volatile molecules that can be easily released to the gas phase by heat or irradiation.

Further studies were devoted to the synthesis of $CO_2$ on carbon foils covered by $H_2O$ ice and triggered by 100 keV protons at 20 and 120 K [257] and covered by $O_2$ ice and triggered by UV irradiation (6.41 eV) at 20 K [258]. The authors concluded that reactions of OH radicals and oxygen atoms with the carbon surface can be a significant source of $CO_2$ in interstellar grains.

Similarly, the synthesis of CO and $CO_2$ on the surface of amorphous carbon grains covered by $H_2O$ and $O_2$ ices under 200 keV proton irradiation at 17 K has been observed [259]. The results reinforced the previous conclusion that the formation of CO and $CO_2$ strongly restricts the lifetime of the solid carbon material in cold astrophysical environments and demonstrated the graphitisation of carbonaceous grains by proton bombardment.

In the same year, Shi *et al.* showed that the synthesis of CO and $CO_2$ on $H_2O$-coated graphite by Lyman-α photons at temperatures from 40 to 120 K should lead to an efficient erosion of the carbon material on the lifetime scales of molecular clouds and protoplanetary disks [260].

Such erosion has also been observed by Chakarov and co-workers during irradiation of $H_2O$ clusters on graphite using wavelengths in the visible spectral range [225,226]. These wavelengths are too long to dissociate $H_2O$ or ionise either $H_2O$ or the graphite and correspond to the peak of the intensity in the interstellar radiation field. It was concluded that sub-threshold hot electrons are implicated in the process reflecting the metallic nature of the graphite substrate. This highlights an important set of processes on conducting carbonaceous materials likely to be ignored if the focus of laboratory study is only on (Lyman-α) UV radiation.

Very recently, UV photochemistry of carbon grains/water ice mixtures has been studied [229]. It was demonstrated there that $CO_2$ can be formed and trapped in carbon/ice mixtures at the temperature of 150 K, which is much higher than its desorption temperature of 85 K. This result may have important implication for our understanding of possible erosion of solids and corresponding chemical and desorption processes in the inner regions of protostellar envelopes and protoplanetary disks.

Increasing molecular complexity in the absence of an ice layer can be demonstrated in the formation of CO, $CO_2$, and formaldehyde ($H_2CO$) on bare amorphous carbon grains by gas-phase O/H atom addition at 10 K as it was shown by Potapov *et al.* [261]. The main result of the study is illustrated by **Figure 21** where difference IR spectra before and after O/H addition of carbon grains are shown. Two bands related to formaldehyde are clearly visible in the spectra.

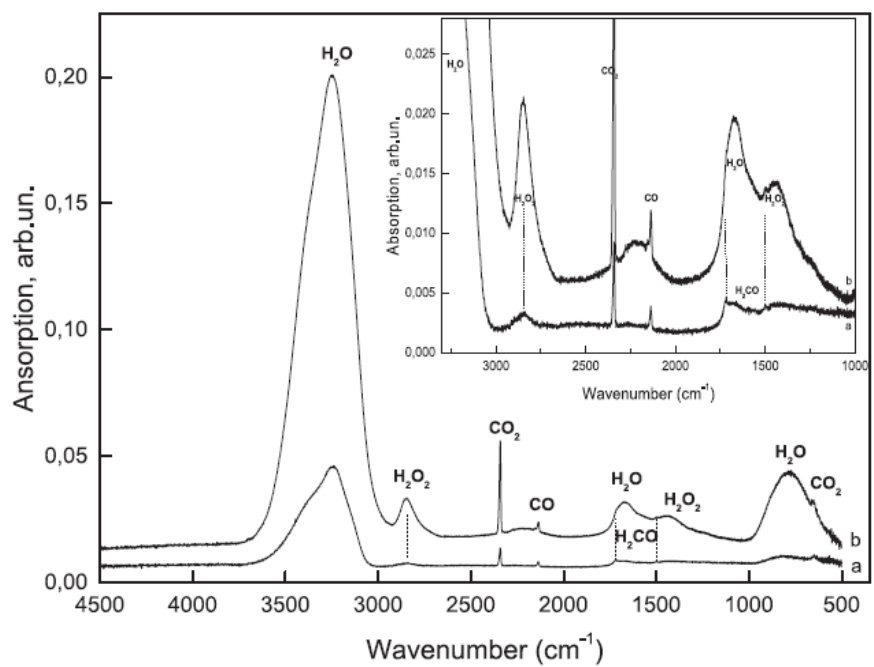

**Figure 21.** Difference spectra before and after O/H bombardment of carbon grains: (a) [$O_2$]/[$H_2$]= 1/60, (b) [$O_2$]/[$H_2$]=10/70. Inset: zoom-in in the 3300 – 1000 cm$^{-1}$ spectral range. Reproduced with permission from [261]. © AAS.

It should be noted that a number of reaction products were formed after O/H bombardment: CO, $CO_2$, $H_2CO$, $H_2O$, and $H_2O_2$. However, the relative amounts of these products depend strongly on the composition of the $O_2/H_2$ mixtures discharged. In the inset in **Figure 21**, which expands the 3300-1000 cm$^{-1}$ spectral range, one can observe that with a low amount of oxygen atoms, mainly CO, $CO_2$, $H_2CO$, and $H_2O$ were formed, while the increase of the oxygen fraction in the $O_2/H_2$ mixture considerably increases the amount of $CO_2$, $H_2O$, and $H_2O_2$, and moderately increases those of CO and $H_2CO$. This is mainly caused by two factors. The increase of the oxygen amount in the $O_2/H_2$ mixture promotes the formation of $H_2O$ and $H_2O_2$. A higher fraction of oxygen promotes the formation of CO via the C+O reaction but also induces the CO-oxidation to form $CO_2$, competing with the CO-hydrogenation.

For the formation of molecules on the surface of grains under O/H atom bombardment the following scenario was proposed in [261]. As discussed by the authors, the hydrogenated fullerene-like carbon grains used in the study have strongly bent graphene layers and many defects that can easily form bonds. Such defects are susceptible to chemisorption of oxygen. In contrast, hydrogen atoms adsorbed on a graphite surface show an adsorption barrier of 0.2 eV. Thus, the first reaction occurring during the H/O bombardment of the grains should involve atomic oxygen:

$$\text{Surface-C} + \text{O} \rightarrow \text{Surface-CO} \tag{34}$$

Surface-CO can be transferred into free CO by further bombardment, so both CO forms can participate in the following reactions with either atomic O or H:

$$\text{CO} + \text{O} \rightarrow \text{CO}_2 \tag{35}$$

$$\text{CO} + \text{H} \rightarrow \text{HCO}; \text{HCO} + \text{H} \rightarrow \text{H}_2\text{CO} \tag{36}$$

$$\text{CO} + \text{H} + \text{H} \rightarrow \text{H}_2\text{CO} \tag{37}$$

The formation of molecules on carbon grains by atom addition can lead to the formation of COMs including prebiotic molecules. The formation of $H_2CO$ is an indication for a possible methanol formation route in such systems (*via* barrier-less hydrogenation) [262,263]:

$$\text{CO} \rightarrow \text{HCO} \rightarrow \text{H}_2\text{CO} \rightarrow \text{H}_3\text{CO} \rightarrow \text{CH}_3\text{OH} \tag{38}$$

(where each arrow represents an H addition). Methanol, in turn, is well-known as a starting point for the formation of more complex organic molecules [264-266]. This new formation route - atom addition to bare carbon grains - could explain the detections of COMs, such as $CH_3OH$ [267], $HC_5N$ and $CH_3CN$ [268] in diffuse and translucent clouds, where dust grains have no or minor ice coating.

H atoms are known to penetrate into $H_2O$ ice only several monolayers [269]. Thus, having a typical view of dust in dense astrophysical environments as a compact core surrounded by a thick ice mantle, molecule formation on the bare grain surfaces by atom addition can only take place during the transition from diffuse to dense cloud and perhaps at the earliest of stages in prestellar core formation. Thereafter "thick" ice mantles are generally assumed and atom addition chemistry occurs on the icy mantle surface. However, if the bare dust material in astrophysical environments is available for the surface processes as proposed in [48,50], bare grain surface processes will occur on the full grain lifecycle from the interstellar, protostellar, and protoplanetary phases.

It is, however, worth remembering that the grain surface will present items as a reactive interface in the presence of photon and charged particle interactions with icy grains. Thus, the processes on the bare grain surface will continue in the presence of suitable radiation.

The penetration depth of UV photons into ice is much greater compared to atoms and was measured to be from several hundred to several thousands of monolayers (ML) for different ices [270,271]. For water ice, 99% of Lyman-$\alpha$ photons are absorbed by $8.9 \times 10^{17}$ molecule cm$^{-2}$ [271]. This is equivalent to 890 ML assuming a monolayer is $10^{15}$ molecule cm$^{-2}$. Thus, ice thickness is not critical for surface molecule formation by UV light. However, the surface area (that defines the ice thickness) is critical with respect to the number of molecules that can be synthesised at the dust-ice interface. If the available grain surface area is much larger than so far believed [50], much more molecules are able to form at the interface.

One more study, which has to be mentioned in this section, was devoted to photochemistry of methanol ice deposited on the surface of amorphous water-rich magnesium silicates containing magnesium carbonates [272]. It was shown that the photolysis of carbonates presented in the substrate was possibly a source of $CO_2$, CO, carbon and oxygen atoms. Organic carbonates can form in the translucent ISM before the formation of molecular ices [273]. However, one more explanation of the increased formation of CO and $CO_2$ on silicates may be ejection of O anions from silicates as was shown for silicates irradiated with low energy electrons [221].

The last but not the least study to be discussed concerned the formation of PAHs on the graphitised surface of silicon carbide SiC grains [274]. It was shown that aromatic species can be efficiently formed on the surface by atomic hydrogen bombardment under physical conditions of the ISM suggesting that top-down routes are important for astrochemistry to explain the abundance of organic species.

### 7.3. Astrochemical Networks

Data relating to thousands of reactions relevant to reaction networks for use in astrochemical models of interstellar environments, circumstellar environments and planetary atmospheres are included in astrochemical databases such as KIDA (http://kida.astrophy.u-bordeaux.fr/) and UDFA (http://udfa.ajmarkwick.net/). As as illustrative example, **Figure 22** shows a gas-phase chemical network driven by $H_3^+$. Apart from illustrating a complexity of astrochemical networks, the figure also provides a link between the gas phase and the solid state as the formation of $H_3^+$ is driven by $H_2$ that is primarily formed on the surface of dust grains.

The availability and reliability of data on molecule formation routes and reaction rates for gas-phase and surface reactions are crucial for the development of reliable gas-grain models. For example, uncertainties in the binding energies in an astrochemical two-phase model of a dark molecular cloud have been shown to have an impact on the outcome of the simulations [275]. Even inclusion of one new reaction between astrophysically relevant species into an astrochemical network may lead to a considerable change of the predicted abundances of many molecules. A good example of this is a recent work on the additon of the $C + H_2 \rightarrow CH_2$ reaction into a well-developed network consisting of 653 species and 7907 reactions [276].

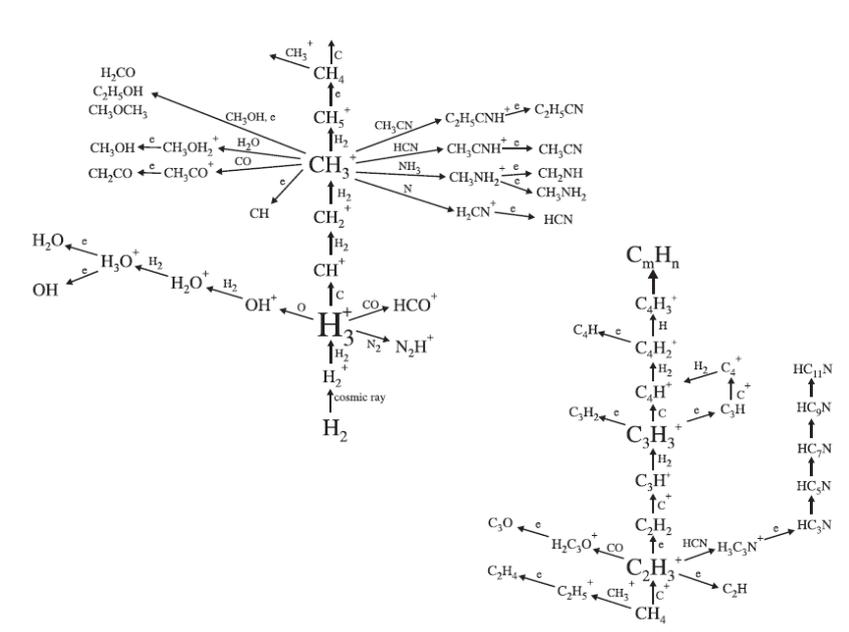

**Figure 22.** Chemical network driven by $H_3^+$ (driven by $H_2$ that is primarily formed on the surface of dust grains). The right hand part connects with its lower end to the top of the left hand part. Reproduced from [277].

As experimentalists, we are far from qualified to discuss the problems of astrochemical models. However, one of the messages that we would like to send to the astrochemical community is that astrochemical modeling should, where appropriate, integrate processes involving dust surfaces. The state-of-the-art of grain surface models has been reviewed and shows a lack of data related to reliable cosmic grain analogues [278,279]. The chemical data discussed in our review, *e.g.,* the formation routes to basic molecules (CO, $CO_2$, $H_2CO$) and the increased reaction rates for the formation of COMs on carbon and silicate grains, should be included into existing models to evaluate their importance for astrochemical networks describing processes taking place in interstellar clouds, protostellar envelopes, and planet-forming disks. Moreover, an understanding of the role of dust in surface reactions can be crucial for simulation of the environmental conditions providing pathways to the formation of prebiotic molecules in the ISM, protostellar envelopes, planet-forming disks and planetary atmospheres.

## 8. Dust/ice Mixtures and Trapping of Water
### 8.1. Dust/ice Mixtures

Experimental studies of dust/ice mixtures are a new direction of research pioneered by one of the authors of this review [49,50,280,281]. Some results, on desorption of $H_2O$ and CO mixed with carbon and silicate grains and on the photochemistry of carbon grains/H$_2$O ice mixtures, have been already discussed in **Sections 6.3** and **7.2.** In this section, we focus on the optical properties of dust/ice mixtures and their astrophysical implication.

Dust grains in cold dense astrophysical environments are typically considered to consist of dust particles and molecular ices. Different grain models are discussed, such as bare dust and ice grains, an ice layer surrounding a compact dust core (both mean physical separation of ice and dust) or ice mixed with individual dust moieties (dust/ice mixtures). Optical properties of grains in different spectral regions are important for the modelling and understanding of the physics in astrophysical environments. The opacity of grains is the base for the estimation of important astrophysical parameters, such as the temperature, structure, and composition of dust, mass loss rates of evolved stars, and the total dust mass. However, different grain models give very different grain parameters (*e.g.*, [282-284]).

To interpret the astronomical spectra of dust grains, spectral data on laboratory dust grain analogues are required. Optical properties of pure water ice, silicate and carbon grains have been intensively investigated in the laboratory (see [281] and references therein). One can try to obtain optical constants of dust/ice mixtures by mixing known pure dust and ice optical constants using a number of mixing rules (effective medium approaches), as has been done in a number of works [285-287]. However, one of the conclusions of the study of the optical constants of silicate grains/water ice mixtures [281] was that differences between measured constants and constants calculated using effective medium approaches show that mathematical mixing (averaging) of the optical constants of water ice and silicates for the determination of the optical properties of silicate/ice mixtures can lead to incorrect results. Thus, in the case of dust/ice mixing in astrophysical environments, to more reliably reproduce observational spectra and to build reliable physico-chemical models of astrophysical environments, one has to use constants (or spectra) measured for physical mixtures of dust and ice.

There is much evidence that ice is mixed with dust in cold astrophysical environments. Analysis of cometary dust particles and laboratory analogues of cosmic dust grains; and evolutionary models of dust grains lead to this conclusion (see [280] and references therein). Recently, the mid-IR spectra of laboratory water ice samples and silicate grains/water ice mixtures were compared with astronomical observations to evaluate the presence of dust/ice mixtures in interstellar environments and circumstellar environments of young stars [280]. While the laboratory measurements of silicate/ice mixtures at 10 K showed no difference in the 3 μm spectral region compared with pure ice measurements — in this case, it is difficult to distinguish between different grain models (ice physically separated from dust or mixed with individual dust moieties) — at higher temperatures, substantial differences occur, leading to substantial differences in mass/temperature estimates. **Figure 23** shows comparisons between the laboratory measurements of silicate grains/water ice mixtures and the observational spectra of a protostar and a protoplanetary disk.

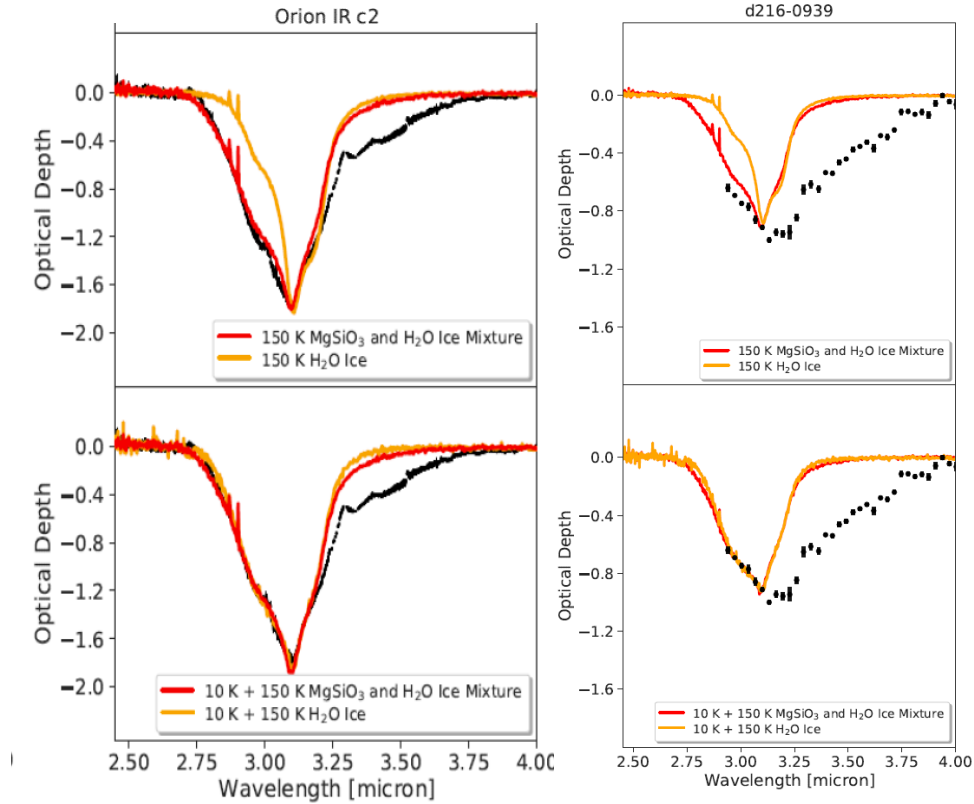

**Figure 23.** Comparison between laboratory measurements and the observational spectra of the protostar Orion IRc2 (left) [288] and the protoplanetary disk d216-0939 (right) [289]. Upper panels: the normalized laboratory spectra of the $H_2O$ (orange) and $MgSiO_3/H_2O$ (red) samples at 150 K compared to the observational spectra (black). Lower panels: the fits of the observational spectra (black) by a mathematical mixing of the $H_2O$ (orange) and $MgSiO_3/H_2O$ (red) 150 and 10 K spectra. The mixing ratios are presented in Table 1. Adapted by the authors from [280].

As can be seen in the upper panels of **Figure 23**, the $MgSiO_3/H_2O$ 150 K measurements show a much broader profile than the pure ice measurements and can explain a substantial part of the blue wing of the observed absorption bands. As the silicate and ice grains in protostellar envelopes and protoplanetary disks will most likely have a broad temperature range depending on their proximity to the star, the observational spectra were also fitted by a mathematical mixing of the laboratory 150 and 10 K spectra to mimic such a temperature distribution of the absorbing material. The results are presented in the lower panels of **Figure 23** and in **Table 1**.

**Table 1.** The mixing ratios for the $H_2O$ and $MgSiO_3/H_2O$ 150 K/10 K spectra and derived mass average temperatures of the protostellar envelopes and protoplanetary disk. Reproduced from [280].

| Source | $H_2O$ | | $MgSiO_3 + H_2O$ | |
|---|---|---|---|---|
| | 150 K/(150 K+10 K) | T / K | 150 K/(150 K+10 K) | T / K |
| Orion BN | 0.165 ± 0.002 | 33.2 ± 0.2 | 0.642 ± 0.006 | 99.9 ± 0.8 |
| Orion IRc2 | 0.236 ± 0.003 | 43.1 ± 0.4 | 1.000 ± 0.016 | 150.0 ± 2.3 |
| Mon R2 IRS3 | 0.133 ± 0.004 | 28.6 ± 0.6 | 0.520 ± 0.011 | 82.8 ± 1.5 |
| d216-0939 | 0.161 ± 0.010 | 32.6 ± 1.4 | 0.565 ± 0.020 | 89.1 ± 2.8 |

From this comparison, it can be clearly seen that the blue wing of the 3 μm ice absorption band can be equally well fitted by mixing the 150 K/10 K spectra of the $H_2O$ or $MgSiO_3/H_2O$ samples. However, the mixing ratios in these two cases differ substantially (**Table 1**). Where the fit by mixing pure $H_2O$ spectra would lead to a conclusion that mainly low-temperature ice (more than 80%) is present in the protostellar envelopes and protoplanetary disk, mixing in the silicate/ice spectra suggests that low-temperature ice accounts for only about 30–50% of the column mass toward the investigated objects. The derived mass-averaged temperatures of the envelope and disk materials in the line of sight differ by a factor of 3, being between 28 and 43 K when using pure water ice and between 83 and 150 K when using the $MgSiO_3/H_2O$ data. The higher values are reasonable for the protostellar envelopes around the three sources investigated and for the protoplanetary disk and show that there is a strong dependence of the derived physical quantities of the absorbing material on the grain model used. The higher temperature values are also supported by the absence of the CO ice feature in the spectra of the envelopes [288] as CO ice desorbs at around 40 K. Thus, the laboratory data can explain the observations assuming reasonable mass-averaged temperatures for the protostellar envelopes and protoplanetary disks demonstrating that a substantial fraction of water ice may be mixed with silicate grains.

### 8.2. Trapping of Water

In laboratory high- and ultrahigh-vacuum experiments, water ice thermally desorbs completely at 160 – 180 K [174,290,291]. However, in the majority of laboratory experiments ice was deposited onto standard laboratory substrates, which are not characteristic of cosmic dust grains. Recently, more realistic laboratory experiments on silicate grains/water ice mixtures by Potapov *et al.* [49,280,281] have demonstrated that a considerable part of water molecules mixed with silicate grains at low temperatures is trapped on silicate grains at temperatures exceeding the desorption temperature of pure water ice. This finding reinforces the results of the calculations on the interaction of water with silicates showing the presence of stronger adsorption sites on the surface, where water molecules may be retained inside the snowline [292,293].

It is known that OH-stretching in hydrated silicates (phyllosilicates) is blue-shifted with respect to pure $H_2O$-ice [294,295]. Spectra of hydrated silicates show a broad absorption from 2.75 to 3.2 μm arising from bound water molecules. This region is the region where the (trapped) water stretching band is observed at 200 K and above. Trapped water probably presents water molecules strongly bound in hydrophilic binding sites on the silicate surface. However, note that silicate/water mixtures at 200 K are not analogues of phyllosilicates where OH groups or $H_2O$ molecules are chemically incorporated into crystalline silicates corresponding to a high-temperature or high-pressure treatment. This suggests that "water reservoirs" in silicates can be formed in interstellar media and circumstellar media of young stars at low temperatures.

Trapped water may be present on silicate grains in the diffuse ISM as well as in planet-forming disks inside the snowline and in planetary atmospheres. Laboratory spectra of silicate grains/water ice mixtures at 200 K were compared with the observations along the sightline towards Cygnus OB2 12, probing the diffuse interstellar medium, done by ISO [288] and Spitzer [296]. The results of this comparison are shown in **Figure 24**. The vertical dashed lines in this figure indicate the observed absorption features of trapped water (2.9 and 6.2 μm). Thus, based on the combination of laboratory data and infrared observations, evidence to suggest the presence of solid-state water in the diffuse ISM has been provided [280].

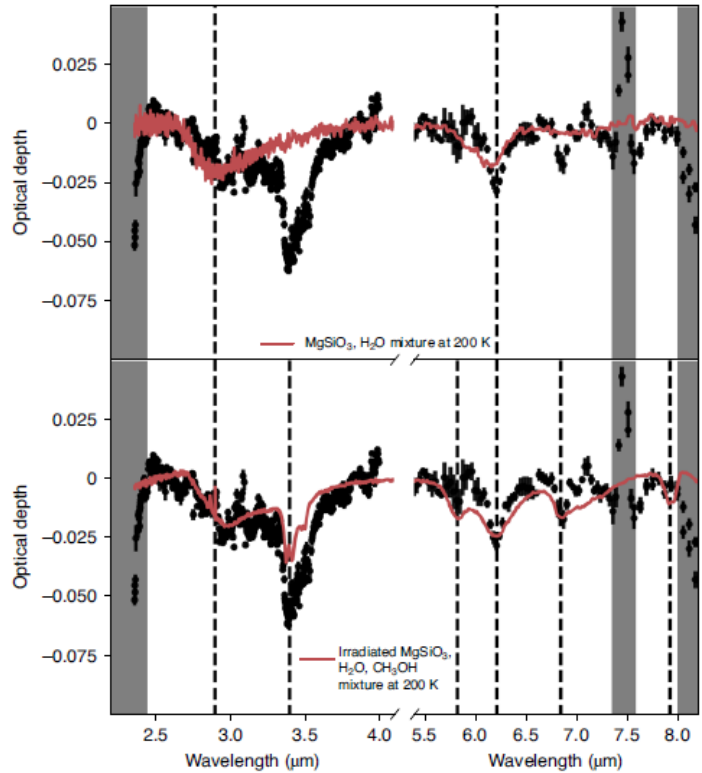

**Figure 24.** Comparisons between the laboratory measurements and the apparent optical depth toward Cyg OB2 12. Top: comparison between the apparent optical depth (black curve) derived from ISO (shorter than 5.0 μm) and Spitzer (longer than 5.5 μm) observations and the laboratory measurements of the mixture of solid-state water and silicates at 200 K (red curve). Bottom: comparison with the laboratory measurements of the ultraviolet-irradiated mixture of solid-state $H_2O$, $CH_3OH$ and silicates warmed to 200 K. The vertical dashed lines indicate the observed absorption features in optical depth profiles. The grey areas indicate the spectral regions dominated by a blend of strong HI emission lines. Reproduced from [280].

Trapped water may survive the transition from cold star-forming regions to planet-forming disks and stay in silicates in the terrestrial planet zone. If trapped water is present on silicate grains in disks inside the snowline, it has two important consequences. First, the efficiency of the formation of rocky and giant planets depends on the amount of solid-state material present at the formation location [297,298]. The enhanced dust densities due to the presence of ice outside of the snowline are assumed to play an important role in the formation of giant planets. In addition, according to coagulation models [85,299] and collision experiments [300], ice-coated grains may stick together much more efficiently, which may lead to the more efficient planetesimals formation. However, recent simulations have shown that the presence of water ice has little impact on dust coagulation in protoplanetary discs, as long as a thick, multilayer ice is not permitted to form [301]. Rocky planets accrete their mass mainly through solid-state material from dust grains, pebbles or planetesimals. Thus, evidence for the presence of solid-state water inside the snowline and an estimation of its abundance will lead to a more complete understanding of the formation process of rocky planets.

Second consequence concerns the origin of water on Earth and terrestrial planets. One of the key aspects to reveal the possibility of life on a rocky planet is to understand the delivery of water to its surface. Assuming that solid-state water exists only beyond the water ice snowline, two scenarios have been proposed. In the so-called dry scenario, the terrestrial planets are initially built up from planetesimals/pebbles inside the snowline with low (or no) water mass fractions and water is delivered to their surfaces by water-rich asteroids formed outside the snowline. Alternatively, in the so-called wet scenario, the planets either accreted a water-rich atmosphere or formed beyond the snowline meaning its time-variable location. However, one more possibility of the wet scenario is that rocky planets formed from planetesimals with water bonded to silicate grains. The water-rich planetesimals option of the wet scenario means that local planetesimals in the time of the Earth formation retained some water at high temperatures through its physisorption or chemisorption on silicate grains [302,303]. Thus, to understand if water can be present in the solid state inside the snowline in protoplanetary disks is the key question for choosing between the different scenarios of the origin of water on Earth and terrestrial planets; and, potentially, on exoplanets.

## 9. Outlook and Future Directions

The results discussed in this review provide deep insight into the physical and chemical processes taking place on the surface of cosmic dust grains. However, there is huge room for future research. In this section, we speculate on some important questions; and outline a few directions and developments related to cosmic dust grain studies, which will help to deepen our understanding of the physics and chemistry of astrophysical environments and even the Origins of Life.

### 9.1. Desorption from Dust Surfaces

Thermal and non-thermal desorption is perhaps the most established on areas of investigations in grain physics and chemistry. However, that does not mean that this topic is sufficiently mature that no further development is necessary. Undoubtedly, in relation to thermal desorption, as we see a rapid increase in the numbers of groups equipped to undertake high quality measurements, it will be necessary that these groups apply consistent methods of analysis, such as discussed in **Section 6.3**. Modellers can then use these data with some guarantee of their quality.

Equally, there are important directions to take in relation to thermal desorption from mixtures of ice and dust. These processes clearly couple the simple two-dimensional desorption that is the current mainstay of such thermal desorption studies with the need to consider three-dimensional diffusion in the bulk ice and three dimensional heat transfer in both the bulk ice and the dust. This may begin to explain some of the discrepancies reported.

As already pointed out several times in this review, the hydrophobic and hydrophilic properties of dust grains may play an important role in the surface processes. TPD of sub-monolayer quantities of $H_2O$ ice seems to be an appropriate method to investigate the presence (and distribution) of hydrophilic binding sites on surfaces (see **Section 6.3**) and should be applied to laboratory analogues of cosmic grains of both, siliceous and carbonaceous origin.

In terms of non-thermal desorption, again the surface science approach to this is well-established. However, there are again obvious directions in which this work could progress. One such direction is in ensuring that both the surface and the gas phase are the subject of studies; ideally in parallel. In this manner, simultaneously using time-of-flight mass spectrometry to probe the gas phase and surface-specific spectroscopies such as FTIR and X-ray photoelectron spectroscopy, it will be possible to establish more completely the inventory of desorbing species and those remaining on the surface during non-thermal processing. This will be particularly important where grain species themselves are intimately involved in the chemistry for example the hot electron chemistry of graphite under illumination with visible light; and the likely chemistry of oxygen anion ($O^-$) released from silicate mineral surfaces *via* electron-stimulated desorption processes initiated by high energy photon and cosmic ray interactions.

There are also opportunities here for computational chemistry. The best quality thermal desorption data, and coverage dependence thereof, are ideal benchmarks for computational investigation. There are also obvious contributions to be made in terms of the branching ratios of desorbates subsequent to electronic excitation of solids.

### 9.2. Diffusion on Dust Surfaces

Diffusion studies on dust surfaces are uncharted territory. Reaction and diffusion are two competing processes and to completely understand how surface reactions proceed, and the catalytic effect of the surface therein, diffusion has to be measured independently. A good example underlining the importance of diffusion measurements has been given in **Section 7.1** concerning the catalytic effect of the dust surface in the formation of ammonium carbamate. Diffusion of several astrophysically relevant molecular species on crystalline water ice [304] and along pores of amorphous water ice [305,306] was measured providing good examples of the studies needed for relevant dust grain analogues. Diffusion rates can be quite different on the surface of carbon and silicate grains particularly when porosity and the presence of surface hydrophilic and hydrophobic binding sites (*e.g.,* silica and silicates are known to have hydrophilic and hydrophobic surface groups [307-309]) are considered.

There is also much opportunity here for computational input. Evaluation of potential energy surfaces of realistic, but simplified, models of grain (and ice) surfaces may yield the information necessary to calculate *ab initio* diffusion rates which might be benchmarked against experimental measurements. Studies on single crystal water ice and silicate surfaces provide this opportunity as they are amenable to investigation by appropriate experimental methods.

## 9.3. Formation of Molecules on Dust Surfaces
### 9.3.1. Formation of Molecules

The results on the formation of $H_2CO$ on the surface of bare carbon grains by atom additions discussed in **Section 7.2** open a door for potential studies. Many examples of possible reactions on bare carbon grains triggered by atom addition leading to astrophysically relevant molecules can be given. As for O/H atom addition, except formaldehyde and methanol, the formation of COMs, *e.g.*, formic acid (HCOOH), glycolaldehyde ($CH_2OHCHO$), and acetaldehyde ($CH_3CHO$), typically synthesized in molecular ices [310,311], may be expected. N/H atom addition to carbon grains may lead to the formation of HCN, methylamine ($CH_3NH_2$), and acetonitrile ($CH_3CN$), potential amino acid precursors in astrophysical environments [29,33,312-314]. N/O atom addition may lead to the formation of the isocyanate radical (NCO) recently detected in the ISM [315].

The study discussed above emphasises a more general point. There is considerable activity associated with atom-grain interactions using atom beam methods. Of course, these represent an important component of the gas phase ISM. However, small radical (*e.g.* OH, SH, CN, $CH_3$, $NH_2$) and carbene/nitrene species (*e.g.* $CH_2$, NH) are found in the gaseous ISM. Studies of their reactions with dust surfaces are therefore clearly warranted. The likely issue is the production of beams of suitable intensity.

Another direction for systematic studies is a continuation of the "carbon grains + $H_2O$ ice + UV" story resulted in the formation of CO and $CO_2$. To make further steps in molecular complexity, one needs to add other simple ice molecules, one by one. The first candidate would be a species containing N to have all four elements necessary for biology (H, C, N, O). Two of such species, abundant in astrophysical environments are $N_2$ and $NH_3$.

In contrast to the extensive knowledge of ion-molecule chemistry in the gas phase, this potentially significant area has received little attention in terms of interactions at grain surfaces. Though focussed on $H_2O$ ice surfaces, the work of Bag *et al.* [316] is illustrative of the potential for unique chemistry associated with such processes. In this work, the formation of $H_2^+$ from $H^+$ on $H_2O$ surfaces was unexpected as was the formation of $D_2^+$ from $D^+$ interactions with both $H_2O$ and $D_2O$ surfaces. The latter was explained by isotope exchange at the $H_2O$ surface prior to reactive formation of $D_2^+$ and the release of an OD radical into the ice to promote further chemistry. There are clearly further opportunities to explore this chemistry which has been emphasised in the recent review by Woon [317]. A logical extension of this work is the application of low energy reactive ion-surface interactions as an analytical tool for investigating molecule formation at dust surfaces [318].

### 9.3.2. Astrocatalysis

In **Section 7.1**, strong evidence of significant catalytic efficiency on the amorphous silicate surface, for instance, Fischer–Tropsch synthesis of $CH_4$ from CO and $H_2$; and Haber–Bosch synthesis of $NH_3$ from $N_2$ and $H_2$, on amorphous silicate surfaces at high temperatures; and the enhanced reaction rate between $NH_3$ and $CO_2$ on amorphous silicate grains at low temperatures were highlighted. When one additionally considers that iron, one of the key transition metals used in industrial catalytic processes, is heavily depleted from the gas phase (into silicate minerals, which nucleate iron nanoparticles during space weathering, and as native iron nanoparticles directly deposited on grain surfaces), nature has provided us with a potential useful catalyst in astrophysical environments. The low-temperature surface conversion of CO to $CO_2$ catalysed by iron was demonstrated [319]. It was shown there that the interaction of Fe and CO leads to the formation of the Fe–CO complex that reacts in the presence of excess CO leading to $CO_2$. It was also proposed that iron could serve as a catalyst for the formation of PAHs and other carbonaceous species in space [320,321]. Other species which have been observed in the ISM such as $TiO_2$, are also known for their catalytic properties (in this case photo-catalytic).

The catalytic formation of COMs is one aspect of astrocatalysis. As those COMs evolve into pre-biotic species then questions begin to arise relating to the potential role of catalytic processes in moving us further towards the complexity of biology. For example, there is a number of important surface thermal reactions for prebiotic astrochemistry, such as $NH_3 + HCN \rightarrow NH_4^+CN^-$ [322], $NH_3 + CH_3CHO \rightarrow NH_2CH(CH_3)OH$ [323], $NH_3 + H_2CO \rightarrow NH_2CH_2OH$ [324], and $CO_2 + 2CH_3NH_2 \rightarrow CH_3NH_3^+CH_3NHCOO^-$ [325]. $NH_4^+CN^-$ can react with $CH_2NH$, a product of the HCN hydrogenation, leading to aminoacetonitrile ($NH_2CH_2CN$) which can, according to the Strecker synthesis, form glycine ($NH_2CH_2COOH$), the simplest amino acid. Aminoalcohol molecules such as α-aminoethanol ($NH_2CH(CH_3)OH$) may be precursors of aminoacids. The reaction of protonated aminomethanol ($NH_2CH_2OH$) with formic acid could lead to protonated glycine, which after dissociative recombination would finally produce glycine [326]. Methylammonium methylcarbamate ($CH_3NH_3^+CH_3NHCOO^-$) acts as a glycine salt precursor in VUV environments. All these reactions may occur on dust grains on much shorter timescales than it is currently thought due to reduction of the reaction barriers as in the case of the $CO_2 + 2NH_3 \rightarrow NH_4^+NH_2COO^-$ reaction discussed in **Section 7.1**. This might also concern hydrogenation reactions such as hydrogenation of CO and CN correspondingly into methanol [262,327] and methylamine [313].

Going further in molecular complexity, is there a role for catalysis in small peptide formation? There is an energy barrier for two amino acids to form a peptide bond. In astrophysical environments this barrier can be lowered by a catalyst, such as the dust grain surface (here again the $CO_2 + 2NH_3$ example is valid). In the chemical laboratory, peptide bond formation, and destruction, is acid-catalysed. This may also be possible in ices presenting mobile protons or on suitable surfaces presenting such species. Alumino-silicate materials can present this type of acid centre and may offer opportunities for catalytic coupling of amino acid. This aligns with the well-known Clay World hypothesis of Cairns-Smith given that clays are hydrothermally-processed alumino-silicate materials [328].

There are therefore significant opportunities to explore astrocatalytic processes both experimentally and computationally *in operando* relevant to a range of environments from the low pressure and low temperature of molecular clouds through the hot and low pressure environments representative of protoplanetary disks to the hot and dense environments that are likely to be observed in exoplanetary atmospheres.

### 9.3.3. Chirality

We should recognise that biology is chiral (handed); the amino acids in terrestrial biology rotate the plane of polarisation of light to the left while the sugars employed rotate the plane of polarisation of light to the right. A fundamental question then arises as to the origin of this homochirality. The thermal and non-thermal chemistry that is currently being explored in relation to the formation of complex molecules on most surfaces and in ices does not produce chirality. Yet, there is strong evidence for an extra-terrestrial origin for the homochirality of the amino acids recovered from the Murchison meteorite [329] and observed in cometary environments [330]. Over many years, several explanations have been put forward to explain this effect. Foremost amongst these is the role played by selective destruction by circularly polarised VUV (CP-VUV) photochemistry [331,332] illustrated in **Figure 25**. However, alternative pathways, also illustrated in **Figure 25**, such as chemistry promoted by spin-polarised hot electrons generated in metal nanoparticles by circularly polarised UV and visible radiation or helical electrons generated by photoemission from ferromagnetic (iron nanoparticles and ionised graphite/graphene) surfaces [333-335] from unpolarised radiation; and synthesis on chiral surfaces have yet to be explored in astrochemistry and astrobiology. The former are especially interesting given that the cross-sections for electron-induced chemistry are significantly higher than for photon-induced chemistry.

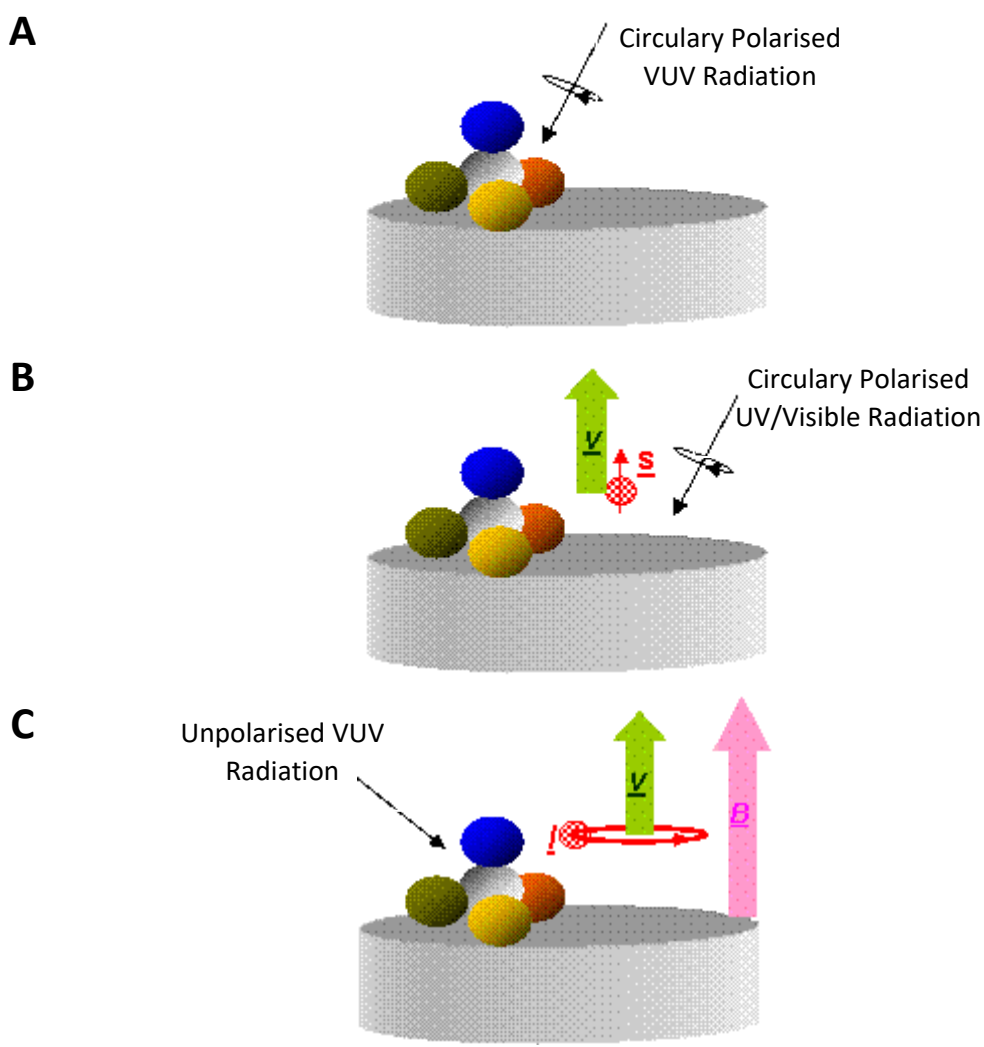

**Figure 25.** Diagrams representing photon-driven enantioselective processing: (A) direct CP-VUV photochemistry; (B) spin-polarised hot electron chemistry initiated by CP-UV/Visible radiation on metallic nanoparticles; and (C) "helical" electron-induced chemistry initiated by unpolarised VUV photoemission in an axial magnetic field.

### 9.3.4. New Experimental Approaches

The two main *in situ* laboratory methods for studying molecules produced in grain surface reactions and in bulk ice reactions are solid-state FTIR spectroscopy and gas-phase mass spectrometry (MS) for detecting directed desorbed products or *via* thermal desorption of the material from the surface. Neither technique is perfect. *In situ* identification of molecules, particularly large molecules, with infrared spectroscopy is problematic. IR spectral bands are weak and broad. Reaction product features often overlap with more intense bands corresponding to reactant species, which make their assignment and subsequent molecular identification difficult and limited. Problems also exist for mass spectrometry as typically applied in laboratory astrophysics experiments. MS is not able to distinguish between molecules of the same mass unless a mass resolution approaching 1 part in $10^6$ is available. Even then, it is generally unable to distinguish isomeric forms of the same molecule. Thus, other techniques should be adopted that can unambiguously identify large molecules newly formed on model cosmic dust and in ice analogues.

One approach potentially leading to the possibility of *in situ* detection and identification of molecules produced on the dust surfaces is a combination of gas-phase high-resolution spectroscopic detection and a low temperature surface desorption experiment. High-resolution laboratory spectra of desorbates taken *in situ* can be directly compared to spectra observed by ground-based radio observatories and give unambiguous information on the molecular composition and structure. In addition, such an approach may provide a means of identifying relevant species produced in surface and solid-state reactions which are not stable at room temperature in the gas phase (*e.g.,* ammonium carbamate, carbamic acid, α-aminoethanol); hence, providing spectral data that will allow their gas-phase detection in interstellar environments and circumstellar environments of young stars. Desorption can be promoted thermally, *cf.* TPD, or using laser-induced heating, *i.e.* laser-induced thermal desorption (LITD) [336]. The latter can achieve local heating of thousands of K s$^{-1}$ by directly exciting the underlying substrate and so promoting rapid, localized desorption without additional reaction. LITD can also be coupled with high-resolution MS to enable mass-based identification of molecular formula [337,338] which undoubtedly would simplify spectroscopic assignment.

There are, to the best of our knowledge, only two very recent examples of gas-phase high-resolution spectroscopy used in studies of cosmic ice analogues. One setup is based on broadband high-resolution THz spectrometer that allowed measurements of spectra of desorbed molecules, $H_2O$ and $CH_3OH$, directly above the ice surface [339]. The second setup was developed using a chirped-pulse Fourier transform microwave spectrometer and benchmarked on the inversion transitions of $NH_3$ desorbed from the walls of a waveguide [340]. Both setups show a potential for COM studies, however, neither of them has been tested on COMs synthesised in surface reactions. Further development of this method is expected and desired.

Two possible routes resolving the issues associated with low- and medium-resolution mass spectrometry techniques in detecting reaction products desorbed from grain surfaces have been identified. The first uses tuneable VUV to selectively ionize different structural isomers [341]. Alternatively, the widely used technique of tandem mass spectrometry maybe applicable in exploring the dissociation of mass-selected ions and hence establishing their structure (*e.g.*, [342]). However, there may be potential issues of sensitivity associated with the latter.

Of course, these techniques can be extended to identifying molecules synthesised thermally or non-thermally in ices on grains. Where such thick films of materials exist, Matrix-Assisted Laser Desorption Ionisation (MALDI) combined with mass spectrometry is possible. This method is widely used in biology and medicine (see, *e.g.,* [343] for a review). However, to the best of our knowledge, there is only one example of its application in astrochemical studies [344]. MALDI using water ice as a matrix, from our point of view, has great potential in astrochemistry and will allow astrochemical studies on the formation of larger prebiotic and biological relevant molecules, *e.g.,* amino acids and peptides. However, the issue of structural isomer identification remains with MALDI.

Identification of species formed by reactions using the techniques discussed above is important and can yield mechanistically useful information including branching ratios where processes can follow two or more simultaneous pathways. Temperature-programmed measurements, in particular, can also be used to a limited extent to extract reaction rates from comparison of modelling with experimental measurements [345]. This is clearly an opportunity for collaboration with modellers in relation to extracting rate parameters for use in astrophysical simulations by modelling experimental measurement. Indeed, Garrod and co-workers have demonstrated the efficacy of this approach [346]. Measurements of isothermal reaction rates can be made using LITD provided that the detection has sufficient sensitivity (*i.e.* using MS). A more flexible approach to isothermal reaction rate measurements on grain surfaces would be possible by adapting approaches developed for studies of semiconductor film growth and catalysis; concentration modulation FTIR spectroscopy for measurements of species on grain surfaces [347,348] and concentration modulation MS for detecting scattered and desorbed gas phase species [349,350].

## 9.4. Dust/Ice Mixtures

The exact nature (including structure and composition) of the solid-state materials in different astrophysical environments can only be determined by spatially resolved multi-wavelength observations over a wide wavelength ranges and by modelling both the thermal emission as well as the absorption towards the central object. New facilities like the James Webb Space Telescope (JWST) are ideally suited to provide us with such data. To fully use these observations, laboratory measurements of reliable analogues of cosmic grains are urgently needed. This provides the motivation for further studying the physico-chemical and spectral properties of dust/ice mixtures (see **Section 8** for more details).

One of the future goals in this direction should be a quantitative study of the influence of different types of silicates as well as of the silicate/ice mass ratio in silicate grains/water ice mixtures on the spectral properties of the mixtures and on the amount and stability of water trapped on silicate grains. A comparison of laboratory and observational spectra will help to reveal the temperature, structure, and composition of silicate grains in astrophysical environments leading to more reliable physico-chemical models of interstellar clouds, protostellar envelopes and planet-forming disks. As for planet-forming disks, information about the structure and composition of dust grains across the snowline will allow better understanding of the process of planet formation including the origin of water on terrestrial planets. Appropriate observations are on the horizon with the launch of the JWST now scheduled for the end of October 2021 and provide a motivation for future collaborative laboratory and observational projects.

## 9.5. Dust Grain Models

Different models describing the structure of dust grains in cold astrophysical environments are discussed in the literature. These include: (i) bare dust and ice grains; (ii) a thick ice layer surrounding a compact dust core; (iii) porous fractal dust core and, possibly, low ice coverage due to the large dust surface area; and (iv) dust/ice mixtures. In the first two models, the most popular for the moment, dust and ice are physically separated and no (or minor) influence of the dust on the surface and bulk processes in ices in interstellar media and circumstellar media of young stars is considered. The latter two models are supported by some results discussed in this review and present a basis for future research.

The experimental results from the Institute of Chemical Sciences of the Heriot-Watt University suggest that, in the process of ice mantle growth on grains, agglomeration of $H_2O$ molecules is possible [47,48]. This may point to the possible availability of the bare grain surface and processes thereon. The experimental results from the Laboratory Astrophysics Group of the Max Planck Institute for Astronomy point to a significantly larger area in grain agglomerates than currently considered. This may result in grains being covered by sub-monolayer or few-layer quantities of ice and, as a consequence, to direct participation of the dust surface in processes in ices in cold cosmic environments [49,50]. The dust/ice mixing model supported by the comparison to observations [280] is in line with these results as the mixing also means the availability of the large surface of grain aggregates to ice molecules and surface processes.

One of the messages of this review is that the role of physico-chemical processes on dust surfaces (adsorption, desorption, mobility, reactivity and formation of species) has to be considered in explaining observations in various astrophysical environments *via* astrochemical networks. The more segregated icy grain models, (i) and (ii), are already included in such networks to some extent. We would suggest that new approaches reflecting models (iii) and (iv) have yet to be developed and, by necessity, should.

## 9.6. Exoplanetary Atmospheres

The study of exoplanetary atmospheres is one of the most exciting and dynamic research directions in astronomy, astrophysics, and astrochemistry. It leads among other intriguing questions to the search for extraterrestrial life and other habitable planets.

Evidence shows that clouds and hazes are common in extrasolar planet and brown dwarf atmospheres (*e.g.*, [351-356]). They play a crucial role in defining temperature profiles and atmospheric dynamics of such bodies. Furthermore, they can create complex organic molecules which are necessary for habitability. Information on the formation, composition, optical properties, and chemistry (homogeneous and heterogeneous) of clouds and hazes is highly desired to analyse observational spectra and to develop reliable atmospheric models. Thus, there is huge room for laboratory studies.

Refractory condensates that form in high-temperature atmospheres of rocky planets and giant gas planets include silicates, such as $SiO_2$, $MgSiO_3$, $Mg_2SiO_4$, $MgFeSiO_4$, $Fe_2SiO_4$; metal oxides, such as $TiO_2$, $Al_2O_3$, $MgO$, $FeO$; and others, for example, $Fe$, $CaTiO_3$, $FeS$ [39].

Silicates are one of the most important classes of atmospheric refractory materials. Nanometre- to micrometre-sized grains of silica, pyroxenes, and olivines were included into the models describing atmospheres of exoplanets and brown dwarfs [351,353,355,357-363]. Evidence of silicate grain absorption and scattering has also been found in their mid-IR spectra [358,364,365] and near-IR spectra [365,366].

Optical properties of solid materials are clearly temperature dependent. Room and low-temperature optical constants have been determined for a large family of species and can be found in the Heidelberg - Jena - St. Petersburg - Database of Optical Constants (http://www.mpia.de/HJPDOC/). However, the absolute majority of known extrasolar planets have temperatures much higher than room temperature: hot Jupiters (T ∼ 1300-3000 K), warm distant gas giants (T ∼ 500-1500 K), hot Neptunes (T ∼ 700-1200 K), and temperate super-Earths (T ∼ 500 K) [367]. There are only a handful of studies of high-temperature optical properties of refractory materials relevant to extrasolar atmospheric clouds and hazes.

Thus, one of the goals of future laboratory studies relevant to exoplanetary atmospheres should be the determination of the optical properties of dust particles (*e.g.,* different types of silicates) at high temperatures in the near- and mid-IR spectral ranges, for which astronomical data are available or will be available soon.

One more goal is to understand, how the condensed phase interacts with the molecular gas (*e.g.,* $H_2$, CO, $H_2O$, $CO_2$, $NH_3$, $CH_4$ and their mixtures) in the absence and presence of energetic fields (*e.g.,* ions, high-energy electrons and UV photons). Sub-questions are grain growth/dissipation, surface chemistry, and depletion of elements. Energetic processing can dramatically change the gas-phase composition of the atmosphere, impacting its radiative properties, thermal structure and dynamics, and can trigger surface reactions on grains leading to the formation of complex organic and prebiotic molecules. There are a number of studies on the gas-phase chemical processes including haze formation at physical conditions relevant to the exoplanetary atmospheres [368-372]. However, gas-grain and grain surface chemistry studies at the high temperatures and pressures relevant to the exoplanetary atmospheres are practically uncharted territory.

## 10. Making the Link to Observations

Astronomical observations, astrochemical modelling, and laboratory experiments should go hand-in-hand in understanding the pathways to the formation of planets, their atmospheres, and to the origins of biology. In the era of new advanced astronomical instruments, such as the Atacama Large Millimetre/submillimetre Array (ALMA), the James Webb Space Telescope (**Figure 26**), and the European Extremely Large Telescope (EELT), comprehensive information about the gas phase and dust grain composition of different astrophysical environments will become available. Laboratory experiments provide spectroscopic fingerprints of species that are necessary for decoding of astronomical spectra and allowing for modelling physico-chemical processes at the conditions of various cosmic environments. Thus, space research must be supported by reliable laboratory data allowing for analysis of observations and for understanding the origin of detected species. Otherwise, the observations are simply stamp-collecting. Below, we provide a few links between the laboratory experiments discussed in this review and observational results "expected" in the nearest future.

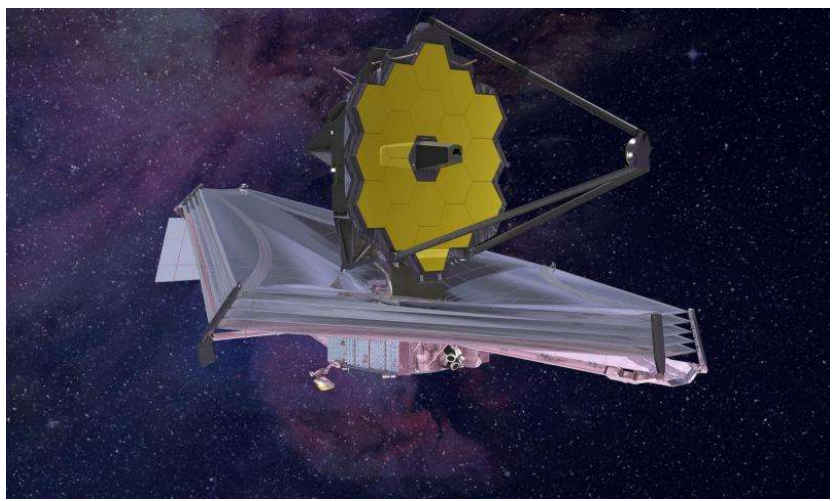

**Figure 26**. The James Webb Space Telescope (JWST). Image credit: Northrop Grumman.

Due to the high sensitivity of modern telescopes, new COMs will be detected in space in the gas phase and perhaps to some extent in the solid state (though the limited resolution makes this difficult). However, their astrochemical origins must be revealed. The studies and methods concerning the surface formation of molecules described in **Sections 6.2, 7.1,** and **7.2** and proposed in **Sections 9.2** and **9.3** will play an important role in respect.

The gas-phase and solid-state abundance of molecular species will be better determined as observation advances. Thus, the link between the gas phase and the solid state provided by adsorption and desorption processes must be understood in detail. This task is relevant to the results and ideas described in Sections **6.1, 6.3,** and **9.1**.

New insights into the origin, structure, and composition of dust grains as described in **Section 4** will not be revealed without the help of the laboratory experiments on dust/ice mixtures described in **Sections 8.1, 8.2, and 9.4** and models discussed in **Section 9.5.**

Finally, the observational studies of exoplanetary atmospheres need a support from the laboratory as discussed in **Section 9.6.**

As a conclusion, we would like to state that strong collaborations between observational, modelling, and experimental groups are needed to ensure a comprehensive view of the physico-chemical processes taking place in space and to reveal the past, present, and future of the Universe.

**Acknowledgments**

AP acknowledges support from the Research Unit FOR 2285 "Debris Disks in Planetary Systems" of the Deutsche Forschungsgemeinschaft (grant JA 2107/3-2). MMcC acknowledges the support of the UK Science and Technology Facilities Council (STFC, ST/M001075/1), the UK Engineering and Physical Science Research Council (EPSRC, EP/D506158/1), and the European Commission Seventh Framework Marie Curie Programme (LASSIE Initial Training Network, grant agreement no. 238258). The authors thank Dr. S. Taj for useful contributions to drafting **Section 4** of this review.